\documentclass[%
 aip,
 sd,%
 amsmath,amssymb,
 reprint,%
twocolumn
]{revtex4-1}
\usepackage[a4paper, paperwidth=18.0cm, paperheight=26.2cm, inner=1.5cm, top=1.5cm, outer=1cm, bottom=1cm, includefoot]{geometry}

\usepackage{amsmath,amssymb}
\usepackage{bm}
\usepackage{graphicx}
\usepackage{caption}
\usepackage{subfigure}
\usepackage{slashbox}
\usepackage[english]{babel}

\usepackage[normalem]{ulem}
\usepackage{color}
\definecolor{epcol}{rgb}{0.0, 0.4, 0}
\definecolor{ascol}{rgb}{0.7,0, 0}
\definecolor{kgcol}{rgb}{0,0, 0.7}

\newcommand{\mA}{{\mathsf A}}
\newcommand{\mB}{{\mathsf B}}

\newcommand{\neiN}{{\mathfrak n}}

\newcommand{\cut}{{\rm cut}}

\newcommand{\lev}{{\rm lev}}

\newcommand{\btheta}{{\bm\theta}}

\usepackage{color}
\definecolor{alertColor}{rgb}{0.7, 0, 0}
\newcommand{\alert}[1]{{\color{alertColor}#1}}

\begin{document}
	
\title{Machine learning of molecular properties: locality and active learning}
\author{Konstantin Gubaev}

\author{Evgeny V. Podryabinkin}

\author{Alexander V. Shapeev}
\affiliation{Skolkovo Institute of Science and Technology, Skolkovo Innovation Center, Nobel str. 3, Moscow, 143026 Russia}

\date{\today}

\begin{abstract}
In recent years the machine learning techniques have shown a great potential in various problems from a multitude of disciplines, including materials design and drug discovery.
The high computational speed on the one hand and the accuracy comparable to that of DFT on another hand make machine learning algorithms efficient for high-throughput screening through chemical and configurational space.
However, the machine learning algorithms available in the literature require large training datasets to reach the chemical accuracy and also show large errors for the so-called outliers---the out-of-sample molecules, not well-represented in the training set.
In the present paper we propose a new machine learning algorithm for predicting molecular properties that addresses these two issues: it is based on a local model of interatomic interactions providing high accuracy when trained on relatively small training sets and an active learning algorithm of optimally choosing the training set that significantly reduces the errors for the outliers.
We compare our model to the other state-of-the-art algorithms from the literature on the widely used benchmark tests.
\end{abstract}


\maketitle

\section{Introduction}
A permanent demand for the discovery of new compounds in numerous fields of industry requires development of the computational tools for prediction of molecular properties. There are many quantum-mechanical algorithms that are able to accurately predict properties of, theoretically, arbitrary atomic systems, however in practice these algorithms are too computationally expensive to be applied to a very large number of molecules.
Density functional theory, which is frequently used due to its favorable trade-off between accuracy and computational cost, is still too time-consuming for high-throughput (rapid) screening of a large amount of molecules.
%or modeling of large systems.

Machine learning (ML) algorithms are a class of algorithms that allow for fast and accurate calculations by constructing a surrogate model which ``interpolates'' between reference quantum-mechanical or experimental data.
In particular, such models were proposed for organic molecules \cite{browning2017genetic,hansen2015machine,ramakrishnan2014quantum,ramakrishnan2015big,ramakrishnan2015machine,huang2016understanding,schutt2017quantum,faber2017fast,gilmer2017neural,huo2017unified} and they can already reach the accuracy of a few kcal/mol or lower by fitting the quantum-mechanical data.
Some of the existing approaches are based on different kernels: kernels based on Coulomb matrix \cite{Rupp2012,rupp2015machine}, the SOAP (smooth overlap of atomic positions) kernel \cite{de2016comparing}, HDAD (histogram of distances, angles and dihedrals \cite{faber2017fast}), BOB (bag of bonds \cite{hansen2015machine}), BAML (bonds, angles, machine learning \cite{huang2016understanding}), MBTR (many-body tensor representations \cite{huo2017unified}). 
Other models are based on neural networks: MPNN (message passing neural networks \cite{gilmer2017neural}), DTNN (deep tensor neural networks \cite{schutt2017quantum}), HIP-NN (Hierarchically Interacting Particle Neural Network \cite{lubbers2017hierarchical}) and SchNet \cite{schutt2017moleculenet}.
 
In this work we propose a new algorithm of fitting of molecular properties.
Our model resembles neural networks in the sense of employing several computing layers.
We show that our model requires less training data to achieve the chemical accuracy when compared against the state-of-the-art approaches on the existing benchmark tests. 
%Furthermore, our active learning algorithms allows us to significantly improve the prediction accuracy for the outliers.
For example, on the benchmark database consisting of 130k molecules \cite{ramakrishnan2014quantum} the majority of recent state-of-the-art algorithms achieve chemical accuracy only when trained on tens of thousands of samples, while our model does it with only few thousands of samples.
We attribute this to our local model of interatomic interaction that effectively relates the molecular properties to the atomic environments and makes predictions for the molecules not present in the training set by accounting for contributions of the individual atomic environments.
%This allows for threating molecules not present in the training set by decomposing them into individual atomic environments and calculating their contributions to the property of interest via some partitioning scheme.

The other problem we address with our algorithm is the issue of the so-called \emph{outliers}.
In the discovery of new molecules the most ``interesting'' molecules are often the most atypical ones.
This is a challenge for ML approaches: if no molecules with similar structure are present in the training set, ML models extrapolate and commit large prediction errors for these outliers.
With the proposed active learning algorithm we reduce the errors for outliers by including the molecules with the most different geometries and compositions in the training set.
This prevents the cases when we are trying to predict properties of molecules that are too different from any of the training samples; instead the properties of out-of-sample molecules are interpolated by the ML model and are thus predicted accurately.

This paper has the following structure: we first formulate the problem of prediction of molecular properties in an ML framework and expose our machine-learning model.
Then, in Section \ref{sec:al} we describe our active learning algorithm.
In Section \ref{sec:results} we compare our algorithm to the existing algorithms \cite{browning2017genetic,hansen2015machine,ramakrishnan2014quantum,ramakrishnan2015big,ramakrishnan2015machine,huang2016understanding,schutt2017quantum,faber2017fast,gilmer2017neural,huo2017unified} on the two benchmark datasets: QM9 \cite{ramakrishnan2014quantum} and QM7 \cite{hansen2015machine}.
The concluding remarks are given in Section \ref{sec:conclusion}.

\section{Machine learning model}\label{sec:alg}
%We formulate the problem as the machine learning task.
We formulate the problem in a machine learning framework in the following way.
Suppose there is a large number, $N$, of molecules whose structure (and composition) is encoded by $x^{(1)},\ldots,x^{(N)}$.
The task is to construct a function $F$ that predicts a certain property (e.g., an atomization energy) of each molecule, $F(x^{(1)}),\ldots,F(x^{(N)})$.
The function $F$, which we refer to as the \emph{model}, will be constructed based on data---the properties of the first $n$ ($n<N$) molecules, $y^{(1)},\ldots,y^{(n)}$, which are called the \emph{labels}.
We will call the set $\{x^{(1)},\ldots,x^{(n)}\}$ together with $\{y^{(1)},\ldots,y^{(n)}\}$ the \emph{labeled dataset}.
The labeled dataset is often chosen randomly.
It is a separate problem to select an optimal subset of molecules such that the prediction error for the rest of the molecules is minimized---we solve it by an \emph{active learning} (or \emph{optimal experiment design}) algorithm outlined in Section \ref{sec:al}.
The remaining set $\{x^{(n+1)},\ldots,x^{(N)}\}$ will be called the \emph{evaluation set} or the \emph{validation set}.
%Selection of the subset $\{x_1,..,x_M\}$ for which to calculate $\{y_i,..,y_M\}$ is a separate and non-trivial task and will be discussed in more details in Section 4. One possible solution is to pick $\{x_1,..,x_M\}$ randomly from $\{x_1,..,x_N\}$.
The model has a number of free parameters $\btheta=(\theta_1,..,\theta_m)$ that are found from minimizing the total loss functional
\begin{equation}\label{loss}
L(\btheta)=\sum_{j=1}^{n} \big(y_j -F\big(\btheta,x^{(j)}\big)\big)^2 \longrightarrow \min. 
\end{equation}
We next describe in detail the model used in the present work and its functional form $F$.
We note that our testing suggests that these models do not need regularization provided that the number of parameters, $m$, is chosen significantly smaller than the number of molecules, $n$.
Hence, $m$ is the only tunable hyperparameter of our model, and we found that a good rule of thumb for choosing it is $n \geqslant 2m$.

%\alert{The number of entities $y^{(n)}$ used for the model training should be larger then the number of free parameters $\btheta$, otherwise the model will be overfitted, learning the non-existing peculiarities of the dataset. As we write further, we can choose among models with different number of parameters, ensuring that the problem \eqref{loss} is overdetermined ($n \geqslant 2m$ is a good rule of thumb) and neither regularization parameters nor early stopping is needed. The only case when this can work bad - if in the dataset there are such labels $m$ and $l$ that $x^{(m)}$ and $x^{(l)}$ are equal or very similar while $y^{(m)}$ and $y^{(l)}$ differ much. In this case the training will be hampered as the first derivatives of the loss functional $F$ w.r.t. parameters $\btheta$  will change dramatically with the small changes of $\btheta$, thus preventing the proper convergence of the optimization algorithm. We expect this effect to occur only if we mix data from different sources, e.g. different levels of ab-initio theory or experiments.} 

We encode a molecule by its atomic positions and types which are denoted collectively by $x$.
Its molecular property (e.g., its atomization energy) is denoted by $y$ and is a function of $x$.
%If for a certain molecule with $N$ atoms we assume an approximation of the explicit form
%$
%y = F(r_1,..,r_N,z_1,..,z_N)\,
%then it should depend on the number of atoms, which substantially limits its transferability, as this number may vary from case to case, each time requiring a new functional form to approximate it.%
To include our knowledge of the underlying physics into the model, we require that the representation of $y$ is invariant with respect to translation, rotation and permutation of chemically equivalent atoms in $x$.

We anticipate that some properties, such as the atomization energy or polarizability, are spatially local, i.e., depend on the local energy or polarizability of interatomic bonds.
Therefore, in addition to the physical invariances, we also introduce locality in our model.

\subsubsection*{Locality}
\begin{figure}
	\centering
	\captionsetup[subfigure]{justification=centering}
	\includegraphics[width=.9\linewidth]{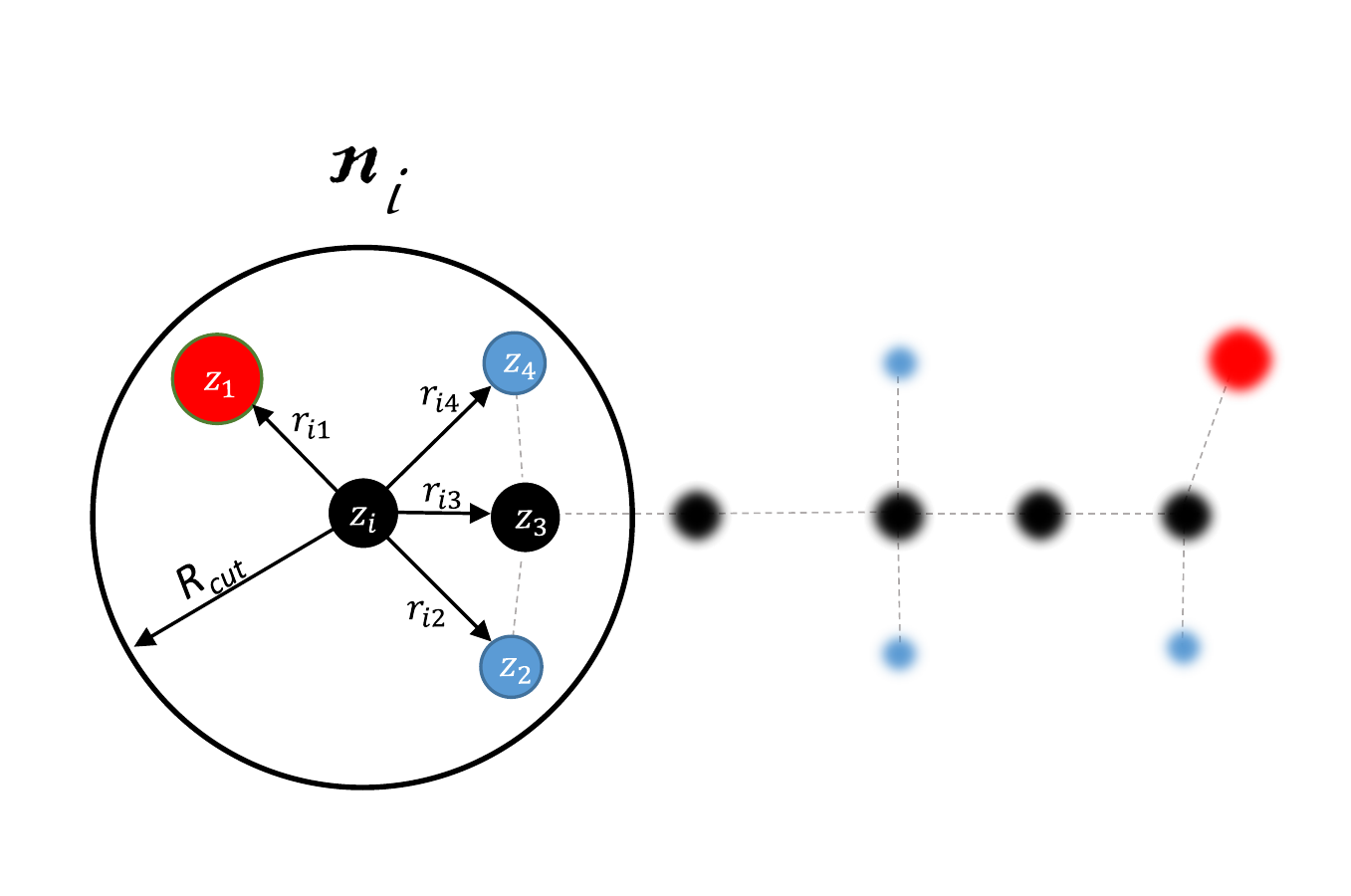}
	\caption {Neighborhood of the $i$-th atom: only the atoms within the cut-off radius $R_\cut$ interact with the central atom}
	\label{fig:interaction-illustration}
\end{figure}

\begin{figure}
	\centering
	\captionsetup[subfigure]{justification=centering}
	\includegraphics[width=.9\linewidth]{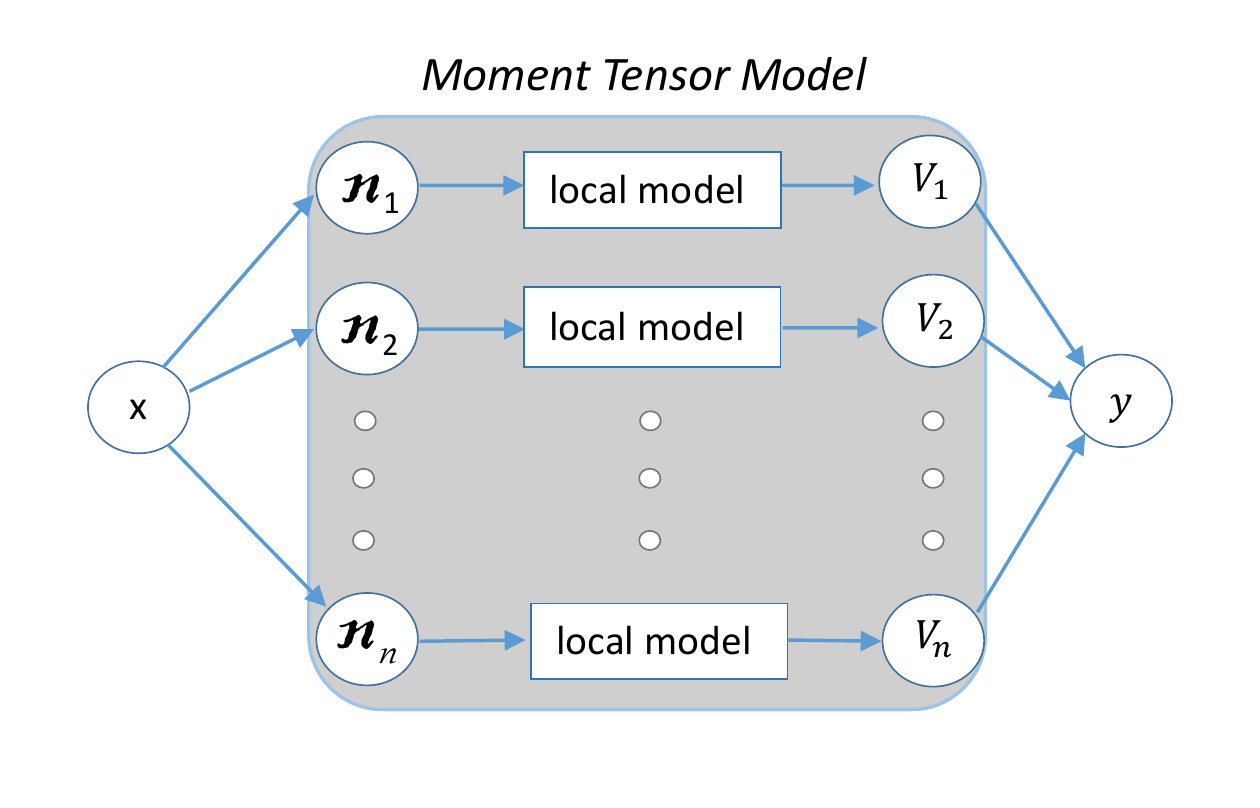}
	\caption {Partitioning scheme: a property $y$ is composed from the contributions $V_i$ of the individual neighborhoods $\neiN_i$}
	\label{fig:common}
\end{figure}

We introduce locality into our model by partitioning $y$ into the sum of contributions of individual atoms. The contribution of atom $i$ depends on its atomic neighborhood: $\neiN_i = \{(r_{ij},z_j)\}$, where $j$ indexes all atoms within the distance $R_{\cut}$ from the atom $i$, and $r_{ij}$ and $z_j$ denote the radius-vectors (i.e., position of the $j$-th atom relative to the $i$-th atom) and the types of the neighboring atoms, respectively; see Figure \ref{fig:interaction-illustration} for an illustration of a neighborhood.
Thus, our model is
\begin{equation}\label{eq:part}
F(x) =  \sum_i V(\neiN_i).
\end{equation}
Here the summation is performed over all the atoms $i$ in the molecule.
Our locality assumption makes this model similar to an \emph{interatomic potential} or a \emph{force field} as used in molecular dynamics.
This model is schematically illustrated in Figure \ref{fig:common}.
The problem thus reduces to finding the right function $V$.

\begin{figure*}
	\centering
	\captionsetup[subfigure]{justification=centering}
	\includegraphics[width=.8\textwidth]{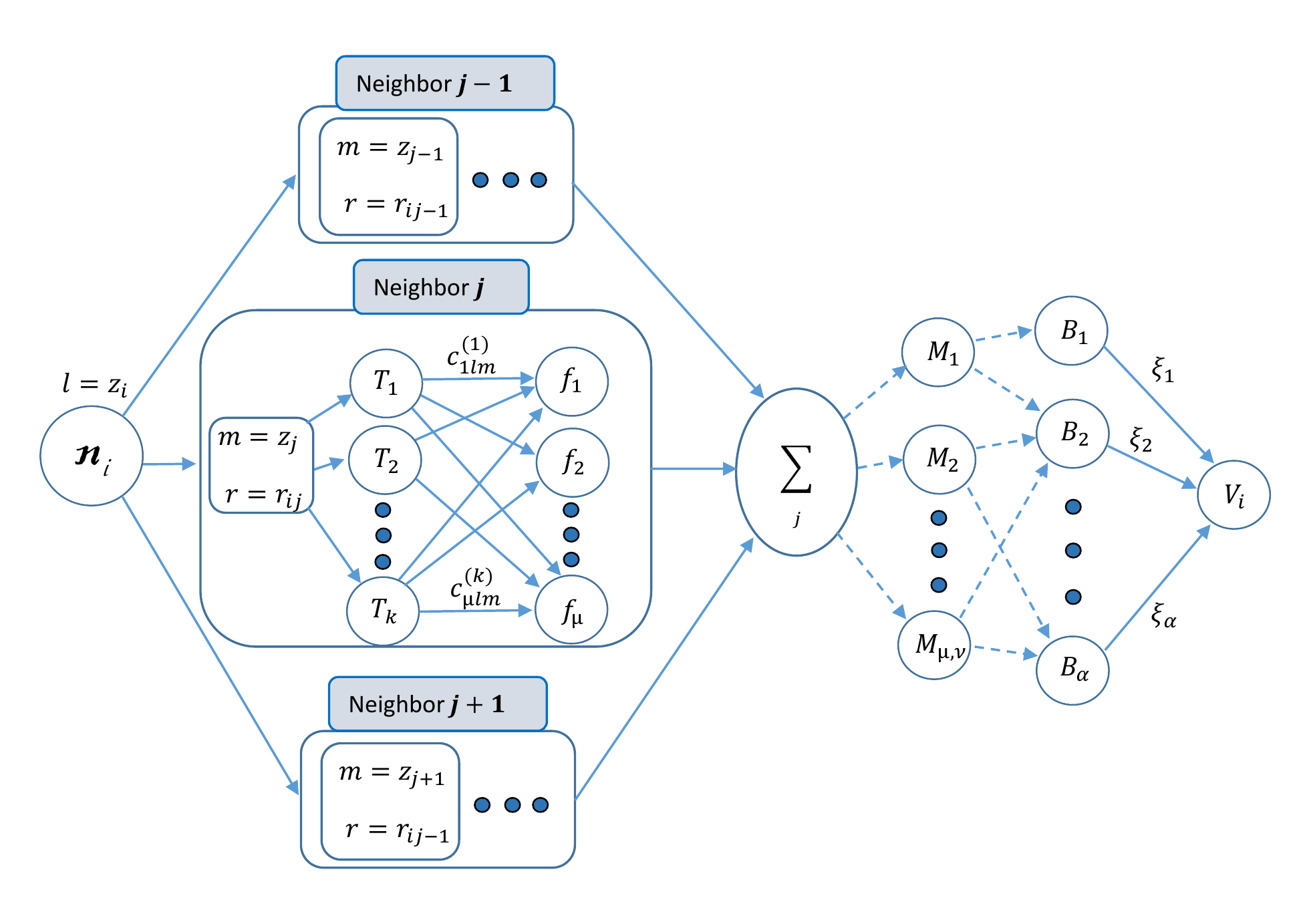}
	\caption {Computation scheme of the moment tensor model (MTM).}
	\label{fig:local}
\end{figure*}

%In this article we will refer to our model in form \eqref{eq:part}, except one case which will be specially discussed. $E = \sum_i V(\neiN_i)$

\subsubsection*{Moment tensors as the descriptors of atomic environment}
We take $V$ as a generalization of the \emph{moment tensor potentials} \cite{shapeev2016moment} to the case of several chemically different atoms.
%It has been proved that the moment tensor potentials are able to describe any local property of a neighborhood with an arbitrary accuracy (at the cost of increasing the number of parameters).
Thus, we let
\begin{equation}\label{lin_exp}
V(\neiN_i)=\sum_{\alpha}\xi_{\alpha}B_{\alpha}(\neiN_i),
\end{equation}
where we call $\xi_{\alpha}$ the \emph{linear parameters}, that are fitted from the data, and $B_{\alpha}(\neiN_i)$ are the basis functions which we describe below.
To define the basis functions, we choose a cut-off radius $R_\cut$ and introduce, as described below, a representation of the neighborhoods which is invariant with respect to rotations and permutations of chemically equivalent atoms.
We note that the translation invariance is already built into \eqref{eq:part}.

We represent atomic environments using the so-called \emph{moment tensor} descriptors of the following form: 
\begin{equation}\label{eq:moments}
M_{\mu,\nu}(\neiN_i)=\sum_{j}f_{\mu}(|r_{ij}|,z_i,z_j)\underbrace {r_{ij}\otimes...\otimes r_{ij}}_\text{$\nu$ times},
\end{equation}
where the index $j$ goes through all the neighbors of atom $i$.
The symbol ``$\otimes$'' stands for the outer product of vectors, thus in \eqref{eq:moments} $r_{ij}\otimes...\otimes r_{ij}$ is the tensor of rank $\nu$.
This way, each descriptor in \eqref{eq:moments} is composed of the radial part $f_{\mu}(|r_{ij}|,z_i,z_j)$ which depends only on the relative distances between atoms and on their chemical types, and the angular part $r_{ij}\otimes...\otimes r_{ij}$ resembling the moments of inertia. Indeed, if in \eqref{eq:moments} we set $f_{\mu}=m^*(z_j)=m_j$, where $m_j$ is the mass of atom of type $j$ and take values of $\nu$ equal to 0, 1 and 2 we will get the descriptors $M_{\mu,0}(\neiN_i)=\sum_{j}m_j$, $M_{\mu,1}(\neiN_i)=\sum_{j}m_j r_{ij}$ and $M_{\mu,2}(\neiN_i)=\sum_{j}m_jr_{ij}\otimes r_{ij}$, having a form resembling the mass, scaled center of the mass, and tensor of inertia of the neighborhood of atom $i$, respectively.
Note that by construction $M_{\mu,\nu}(\neiN_i)$ are invariant with respect to rotations (in a tensorial sense) and permutations of chemically equivalent atoms.
It should be emphasized that $M_{\mu,0}$ are the standard two-body descriptors of atomic environments that do not contain information about angles between bonds.
The general moment tensor descriptors $M_{\mu,\nu}$ remain two-body and thus offer an alternative way of including the angular information---the traditional way is to include at least three-body descriptors.

The descriptors $M_{\mu,\nu}$ defined above are tensors of different ranks.
We hence define the basis functions as all possible contractions of these tensors to a scalar, for example:
\begin{align*}
B_0(\neiN_i) &= M_{0,0}(\neiN_i),
\\
B_1(\neiN_i) &= M_{0,1}(\neiN_i)\cdot M_{0,1}(\neiN_i),
\\
B_2(\neiN_i) &= M_{0,0}(\neiN_i)(M_{0,2}(\neiN_i):M_{0,2}(\neiN_i)).
\end{align*}
Note that all the functions $B_{\alpha}(\neiN_i)$ constructed this way are rotation invariant.
The basis functions $B_\alpha$ resemble invariant polynomials of the atomic positions $r_{ij}$ (if the radial functions $f_{\mu}(|r_{ij}|, z_i, z_j)$ were polynomials of $r_{ij}$ then $B_\alpha$ would have been polynomials).
We therefore define a \alert{level} of $M_{\mu,\nu}$ by $\alert{\lev} M_{\mu,\nu} = 2\mu+\nu$ and if $B_\alpha$ is obtained from $M_{\mu_1,\nu_1}$, $M_{\mu_2,\nu_2}$, \ldots, then $\alert{\lev} B_\alpha = (2\mu_1+\nu_1) + (2\mu_2+\nu_2) + \ldots$.
In practice we obtain different models by including all basis functions such that $\alert{\lev} B_\alpha \leq d$.
We then denote such a model by MTM$_d$
(moment tensor model of \alert{level} $d$).
The larger the \alert{level} is, the more parameters $\xi_{\alpha}$ it has.

The radial functions $f_{\mu}$ from \eqref{eq:moments} are represented in the following way:
\begin{align}\label{rad}
f_{\mu}(r, z_i, z_j) &\phantom{:}= \sum_k c^{(k)}_{\mu, z_i, z_j}Q^{(k)}(r),
\qquad\text{where}\\\notag
Q^{(k)}(r) &:= T_k(r)(R_{\cut} - r)^2.
\end{align}
Here $T_k(r)$ are the Chebyshev polynomials on the interval $[R_{\min},R_{\cut}]$ and the term $(R_{\cut}-r)^2$ is introduced to ensure a smooth cut-off to 0. Taking into account that in real atomic systems atoms never stay too close to each other, we can always, in practice, choose some minimal distance $R_{\min}$. 

The point that makes this model different from the single-component moment tensor potentials \cite{shapeev2016moment} is that now there is a dependence of the radial functions $f_{\mu}$ on the type of the central atom $z_i$ and the types of the other atoms in the neighborhood $z_j$.
This dependence is encoded in the $c^{(k)}_{\mu, z_i, z_j}$ coefficients.
Together with $\xi_{\alpha}$ from \eqref{lin_exp} they form the set of model parameters $\btheta = \big(\xi_{\alpha}, c^{(k)}_{\mu,z_i,z_j}\big)$ that are found during the fitting procedure, see Figure \ref{fig:local}. 

For the case of single-component systems it was proved in Ref.\ \onlinecite{shapeev2016moment} that the moment tensor descriptors form a complete set of descriptors, in the sense that any function of a local atomistic environment can be approximated by a polynomial of the descriptors.
It is hence easy to see that the same property holds for the multi-component systems: since the radial functions $f_{\mu}$ have an arbitrary dependence on the types of the neighboring atoms, the moments $M_{\mu,\nu}$ can describe the way atoms of type one, two, etc., are placed around the $i$-th atom independently of the other types of atoms.
Since the description of each atomic type in the neighborhood is complete, the description of the entire neighborhood is complete.

%\commentas{TODO: revise}
%\commentkg{systematically improvable}
%Summing this section up, we proposed a family of \emph{moment tensor models} (MTMs) for the fitting of the molecular properties, which assume locality of interatomic interactions and use partitioning scheme \eqref{eq:part}. To represent the atomic environments we use Moment Tensor descriptors \eqref{eq:moments}. Following the Theorem 2.1 from \cite{shapeev2016moment} which can be generalized to the multiple species case, the basis $B_{\alpha}$ can approximate any \commentkg{rotationally and permutationally invariant polynomial function of vector variables in euclidean space? Then how to connect it with PES?} function. This means that by increasing the number of distinct values for $\mu$ and $\nu$ present in the current model and hence, by getting more basis functions $B_{\alpha}$, we can improve the accuracy of the model to the required extent, so we say that MTMs are \emph{systematically improvable}. Extending the range for parameters $\mu$ and $\nu$ means including representations of more complex many-body interactions. 
%This increases the computational time consumption but potentially allows achieving higher accuracy. \commentas{TODO: explain (define) what are 16, 20, 24, etc.}MTM$_{16}$, MTM$_{20}$ and MTM$_{24}$ models used in the present work have different amount of parameters and thus behave differently, see Figure \ref{fig:MAE_random}.

\subsubsection*{Intensive quantities}

To further include the knowledge of physics/chemistry into our model, we distinguish between intensive and extensive quantities.
An extensive quantity (such as the atomization energy) scales with the number of atoms, while an intensive quantity (such as the HOMO/LUMO level) remains of the same order when the number of atoms in a molecule increases.
The above description of the model is valid for extensive quantities, while for an intensive quantity we modify the partition relation \eqref{eq:part} as
\begin{equation}\label{eq:part:intensive}
F(x) =  \frac{1}{\#(x)} \sum_i V(\neiN_i),
\end{equation}
where $\#(x)$ is the number of atoms in $x$.

\subsubsection*{A nonlocal model}

To go beyond the locality assumption \eqref{eq:part}, we include nonlocal effects by introducing two different local models $V_1$ and $V_2$ (each with its own set of parameters) and let
\begin{align*}
v_1 &= \sum_i V_1(\neiN_i),
\qquad\text{and}
\\
v_2 &= \sum_i V_2(\neiN_i).
\end{align*}
We then define the nonlocal model \emph{nlMTM} in the following form
\begin{align} \notag
F_{\rm nl}(x) &=  p_1v_1+p_2v_2+p_3v^2_1+p_4v_2^2+p_5v_1v_2
\\&\quad +p_6v_1^3+p_7v_2^3+p_8v_1^2v_2+p_9v_1v_2^2.
\label{eq:nonlocal-model}
\end{align}

\begin{figure*}
	\centering
	\captionsetup[subfigure]{justification=centering}
	\includegraphics[width=.8\textwidth]{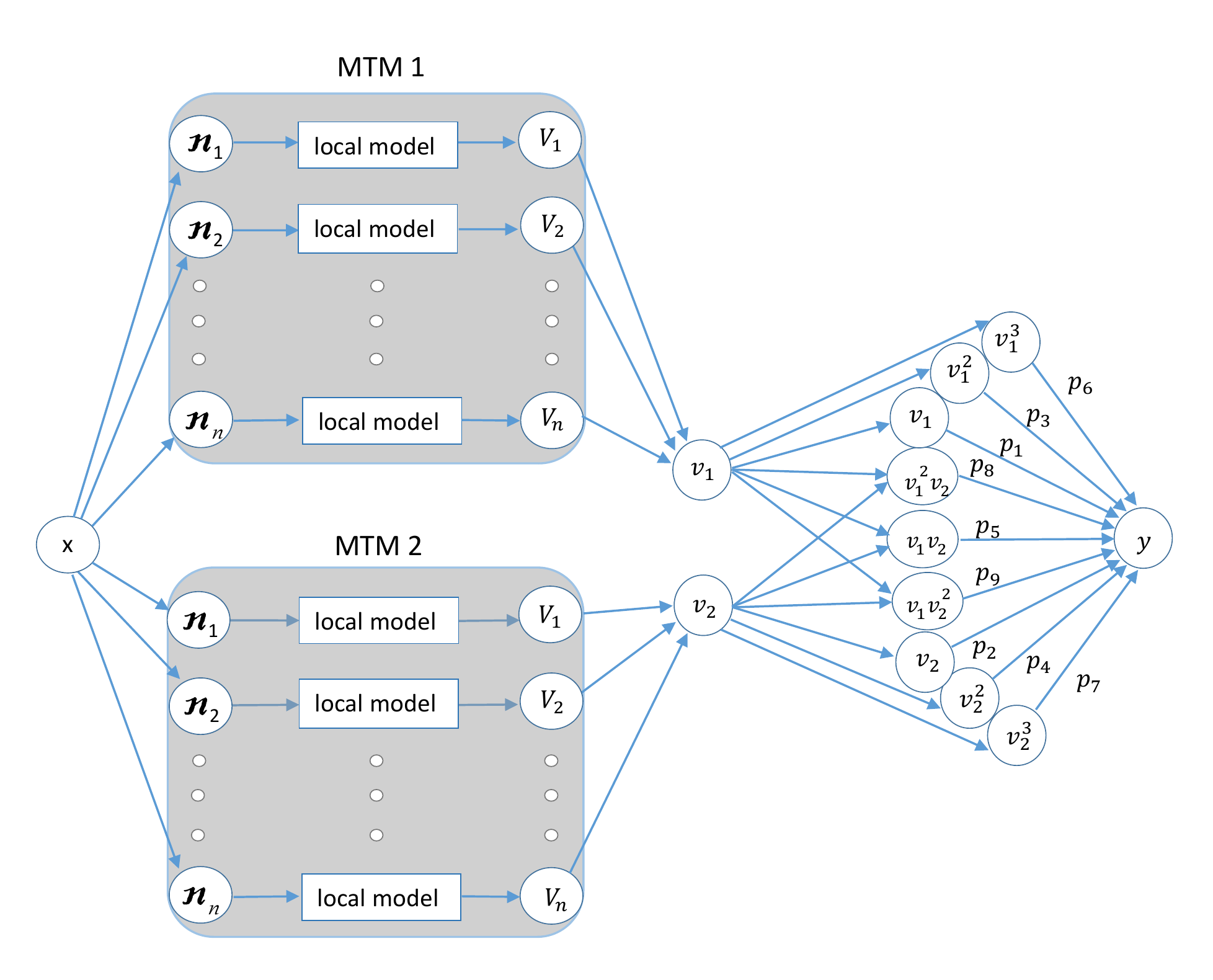}
	\caption {The nonlocal model, nlMTM, constructed from the two local models. $p_1,\ldots,p_9$ are the additional fitting parameters.}
	\label{fig:noloc}
\end{figure*}

Figure \ref{fig:noloc} shows how the parameters $p_i$ form the additional computing ``layer'', if compared to the local model in Figure \ref{fig:common} and extend the set of parameters, $\btheta = \big(\xi_{\alpha}, c^{(k)}_{\mu,z_i,z_j}, p_i\big)$, that are found in the training process.
Such a model architecture is motivated by the fact that molecular orbitals depend largely on local environments described by several $V_i$ (although we only considered $i=1,2$).
However, these orbitals get occupied by electrons not independently of each other, hence we assume a nonlinear dependence of the answer on the local features \eqref{eq:nonlocal-model}.

\section {Active Learning}\label{sec:al}

The accuracy of prediction depends not only on the machine learning model but also on the training set that is used for the fitting.
On the one hand, the size of the training set is typically limited by the amount of experiments or ab initio calculations we can conduct. On the other hand, the training set should represent the full variety of molecules to prevent extrapolation while the evaluation of molecular properties.
It has been shown \cite{Huang2017opt} that an optimal choice of the training set of a fixed size can, in principle, significantly reduce the out-of-sample errors, if we were allowed to use the labels ($y^{(i)}$) of all the available data.
However, a practical selection algorithm needs to choose the molecules for the training set based only on the unlabeled data ($x^{(i)}$), since in practice we want to compute the labels only after selection.
The approaches related to the construction (or selection) of an optimal training set are known as \emph{active learning} approaches.
%Active learning can be considered as selecting the samples from the dataset for learning (preliminary labeling them if selected samples has no labels). 

%We also refer to a recent work qwe

Recently an approach for active learning of interatomic potentials with a linear dependence on the model parameters has been proposed \cite{podryabinkin2016active}. This approach is based on a D-optimality criterion for selecting the training dataset, which is equivalent to choosing atomistic configurations maximizing the determinant of the matrix of the linear equations on the model parameters.
This algorithm effectively detects samples on which the model extrapolates.
Hence, training on such samples prevents extrapolation and thus ensures reliable treatment of the remaining molecules at the evaluation stage.

\subsection{Generalized D-optimality criterion}
 
The model proposed in this paper has a nonlinear dependence on its parameters $\btheta$.
Therefore we propose a generalization of the D-optimality criterion to the nonlinear case.
To that end, we assume that the values of the parameters $\bar{\btheta}$, which are found from the training procedure (Section \ref{sec:alg}) are already near the optimal ones and we hence linearize each term in the loss function \eqref{loss} with respect to the parameters:
 \[
 	 y^{(i)} - F\big(\btheta,x^{(i)}\big) \approx y^{(i)} - \sum_{j} (\theta_j-\bar{\theta}_j) \, \frac{\partial F}{\partial \theta_j}\big(\bar{\btheta},x^{(i)}\big).
 \]
We can then interpret the fitting as the solution of the following overdetermined system of equations with respect to $\theta_j$:
\[
	\sum_{j=1}^m \theta_j \frac{\partial F}{\partial \theta_j}\big(\bar{\btheta},x^{(i)}\big) = y^{(i)} + \bar{\theta}_j \frac{\partial F}{\partial \theta_j}\big(\bar{\btheta},x^{(i)}\big),
\]
\[
 	\quad i = 1, \ldots, n,
\]
where $n$ is the size of the labeled dataset $\{x^{(1)}, \ldots, x^{(n)}\}$. 
The matrix of this system is a tall Jacobi matrix
 \[
\mB = \begin{pmatrix}
\frac{\partial F}{\partial \theta_1}\big(\bar{\btheta},x^{(1)}\big) & \ldots & \frac{\partial F}{\partial \theta_m}\big(\bar{\btheta},x^{(1)}\big) \\
\vdots & \ddots & \vdots \\
\frac{\partial F}{\partial \theta_1}\big(\bar{\btheta},x^{(n)}\big) & \ldots & \frac{\partial F}{\partial \theta_m}\big(\bar{\btheta},x^{(n)}\big) \\
\end{pmatrix},
 \]
where each row corresponds to a particular molecule from the training set.

Next, we select for training a subset of molecules yielding the most linearly independent rows in $B$. This is equivalent to finding a square $m\times m$ submatrix $\mA$ of maximal volume (i.e., with maximal value of $|{\rm det}(\mA)|$).
We do it by using the so-called maxvol algorithm \cite{oseledets2010how-to1770566}.

Active learning allows us not only to extract a D-optimal training set from the labeled dataset, but also (probably, equally importantly) to estimate, at the evaluation stage, the novelty of molecules compared to those in the training set.
If some molecule $x^*$ significantly differs from all other molecules in the training set then its properties cannot be reliably calculated.
Therefore it should be labeled and added to the training set instead. 

We define the ``novelty'' grade $\gamma(x^*)$ as the maximal factor by which $|{\rm det}(\mA)|$ can grow if $x^*$ is added to the training set.
According to Ref.\ \onlinecite{oseledets2010how-to1770566} it can be calculated as 
\[
 %	\label{eq:alcrit}
 	\gamma(x^*) = \max_{1\leq j \leq n} (|c_j|),
\]
where 
\[
 c = \bigg(
 	\frac{\partial F}{\partial \theta_1}\big(\bar{\btheta},x^*\big) \ldots \frac{\partial F}{\partial \theta_n}\big(\bar{\btheta},x^*\big)
 \bigg) \mA^{-1} =: b^* \mA^{-1}.
\]
%As shown in \cite{oseledets2010how-to1770566}, 
%swapping of $j^*$-th row of $\mA$, where $j^* = \arg\max_{1\leq j \leq n} (c_j)$, and the row $b^*$ leads to $\gamma(x^*)$ times increase of $|{\rm det}(\mA)|$. 
Thus, we add the molecule $x^*$ to the training set if $\gamma(x^*) \geq \gamma_{\rm trsh}$, where $\gamma_{\rm trsh} \geq 1$ is a threshold parameter that prevents the algorithm from training on the molecules with not sufficiently high novelty grade $\gamma$.

Taking into account that $F(x) =  \sum_i V(\neiN_i)$, for a certain configuration $x^{(n)}$ one can write: 
\[
\frac{\partial F}{\partial \theta_m}\big(\bar{\btheta},x^*\big)=\sum_i\frac{\partial V(\neiN_i)}{\partial \theta_m}\big(\bar{\btheta},x^*\big)
,
\]
where individual atomic environments of $x^*$ are enumerated with $i$.
Such partitioning over neighborhoods prevents an AL algorithm from selecting molecules composed from already known neighborhoods, as for such molecule the vectors $\frac{\partial V(\neiN_i)}{\partial \theta_m}$ will be linearly dependent on such vectors of smaller molecules and thus the big molecule will not be recognized as sufficiently new one to be chosen.

\subsection{Active learning scheme}\label{sec:al-scheme}

We next describe our active learning procedure.
It expands the training set iteratively, each time increasing its size by no more than 10\%.

 \begin{enumerate}
 	\item[0.]
 	Start with a random initial training set.
 	\item[1.]
 	Train the model on the current training set.
 	% and check its accuracy on the validation set.
 	%Following our objective, it contains all molecules except those from the training set. If some molecule enters the training set, it is excluded from the validation set.
 	\item[2.]
 	Using the active learning algorithm \cite{podryabinkin2016active} select molecules with $\gamma \geq \gamma_{\rm trsh}$, and add either all such molecules or the molecules with the maximal $\gamma$, such that the size of the training set increases by 10\% or less.
 	
 	\item[3.] 
 	Unless satisfied with the current model (see the discussion below), go to step 1.
 \end{enumerate}

%As the first step, we randomly select some molecules to form the initial training set. The required amount depends on the number of parameters in the model. For example, if the model has 1500 parameters to be found, training on 1000 molecules will lead to some overfitting, reducing the accuracy on the validation set. On the other hand, due to iterative nature of this procedure and expanding of training set, sooner or later the required amount will be got, as the biggest number of parameters among models we used does not exceed 2100.
 
% Then below cycle is repeated until required conditions are met, which is discussed further:

For the purpose of studying the performance of the proposed algorithm, we stopped the loop (as mentioned in step 3) when the number of molecules in the training set reached $6000$.
However, in practice, other stopping criterion could be considered:
 %\begin{itemize}
 %	\item 
 the difference in accuracy of the models on two consecutive iterations on the selected (at step 2) molecules is too small.
 	This is an intuitive criterion, however we have observed that sometimes the accuracy improvement on one particular iteration may be small, while on the next one it can increase again.
 	Therefore, it may be better to track improvement on several iterations of the cycle.

\section{Results}\label{sec:results}

We have conducted a number of tests in order to clarify the following three questions: what accuracy can our model achieve on a dataset when trained on its randomly chosen subset, how the accuracy can be improved using our active learning technique (Section \ref{sec:al}), and how many training samples are required to reach the chemical accuracy, 1 kcal/mol.
We remind that we use the following terminology: the \emph{training set} is the set on which we train our model and the \emph{validation set} consists of the full database excluding the training set. All errors quoted below are measured on the validation set. 

%We have investigated the influence of maximal $\mu$ from \eqref{eq:moments} and $k$ from \eqref{rad} on the fitting accuracy. We found that in practice there is no need to take $\mu$ greater then six and $k$ greater then eight, as exceeding this limits provides only slight improvement of accuracy.
We have tested the models of \alert{level} 16, 20 and 24 denoted by MTM$_{16}$, MTM$_{20}$, MTM$_{24}$.
The fitting of the models was done with the Broyden-Fletcher-Goldfarb-Shanno (BFGS) optimization algorithm, by performing between 2000 and 5000 iterations.

\subsection*{Fitting enthalpy on QM9}

First, we fit our model on the so-called QM9 dataset \cite{ramakrishnan2014quantum} consisting of 130831 molecules formed by C,H,O,N,F atoms with up to nine heavy (C,O,N,F) atoms. This is a subset of the originally proposed database \cite{ruddigkeit2012enumeration} consisting of 166 billion of organic molecules.
The number 130831 excludes 3054 molecules from the database that failed a consistency test, as reported in Ref.\ \onlinecite{ramakrishnan2014quantum}.
Following the existing works \cite{ramakrishnan2015big, ramakrishnan2015machine,schutt2017quantum,gilmer2017neural,huo2017unified} we demonstrate the performance of our method by fitting the enthalpy (or atomization energy) at 300 K.

\subsubsection*{Random choice of the training dataset}

\begin{figure}[h!]
	\centering
	\captionsetup[subfigure]{justification=centering}
	\includegraphics[scale=.65]{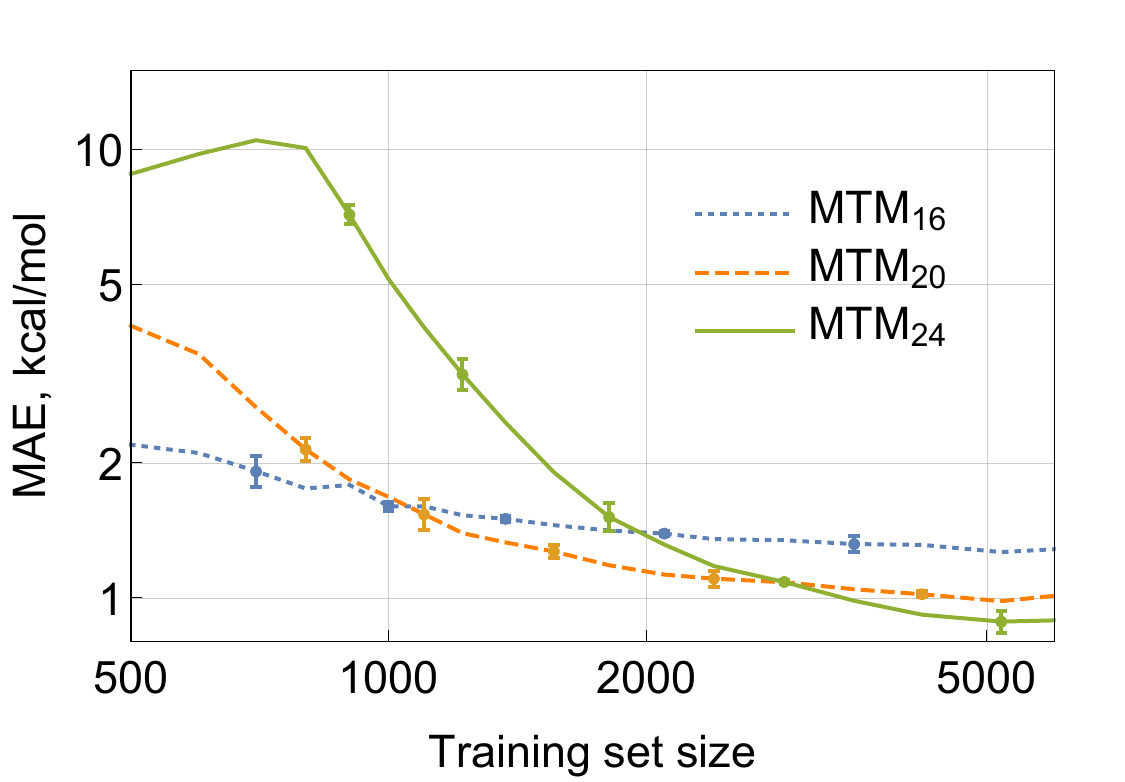}
	\caption {Random choice of the training set: dependence of MAE on the training set size. Different curves show different models: the higher is the number, the more parameters the model has.}
	\label{fig:MAE_random}
\end{figure}

To investigate the accuracy of the MTMs with different number of parameters we have calculated the learning curves showing the dependence of the mean absolute error (MAE) on the training set size, see Figure \ref{fig:MAE_random}.
Our results are averaged over three independent random choices of the training set.
As expected, the models with fewer parameters show good results for small training datasets, but are outperformed by the models with more parameters as the number of training samples grows.

We next compare different models by how fast (i.e., with what training dataset size) they reach chemical accuracy.
Table \ref{tab:gdb9comp} lists the prediction errors for the MTMs and the existing state-of-the-art methods when training on random training sets.
While filling this table, we used MTM$_{16}$ for the 1k training set size, MTM$_{20}$ for the 3.5k training set size, and MTM$_{24}$ for the 10k training set size. For the sizes of 25k and 50k we used the MTM$_{28}$ model which has more parameters to fit then MTM$_{24}$. At the same time, from Fig.\ \ref{fig:MAE_random} it can be seen that using either MTM$_{20}$ or MTM$_{24}$ models still provides competitive results.

The aSLATM \cite{Huang2017opt} model reaches the chemical accuracy with the training set size about 3200, while the learning curve from Figure \ref{fig:MAE_random} shows that MTM$_{24}$ model requires almost the same number, 3500 molecules to reach such accuracy.
We obtained the results for MTM$_{16}$, MTM$_{20}$, and MTM$_{24}$ by doing 2000 iterations of the BFGS algorithm and 5000 iterations for MTM$_{28}$.
For a more accurate comparison with aSLATM, we trained MTM$_{24}$ model on 10 different random samples of training set with 3000 molecules by doing 3000 iterations of BFGS.
This provided us with a validation MAE of $1.006\pm0.0316$ kcal/mol.

As can be seen from Table \ref{tab:gdb9comp}, for training set sizes not more than 10k MTM shows a better learning curve than the existing state-of-the-art methods and only the model from the Ref.\ \onlinecite{Huang2017opt} shows nearly equal results.
The works using deep NNs (Refs.\ \onlinecite{schutt2017moleculenet} and \onlinecite{lubbers2017hierarchical}) show better accuracy for training set sizes over 50k.
Ref.\ \onlinecite{schutt2017quantum} and \onlinecite{faber2017fast} need 25k and 35k training samples to attain the chemical accuracy, respectively.
%In the article \cite{faber2017fast} describing a lot of machine learning models, the best one reached chemical accuracy on training set size of approx. 36k.
We note of another work \cite{gilmer2017neural} that also reaches the chemical accuracy, but it only reports MAE of 0.55 kcal/mol while training on 110k molecules plus another 10k molecules used as a hold-out set for the early stopping criterion. 

\begin{table}
\begin{tabular}{|l|c|c|c|c|c|c|c|}
	\hline
	 & \multicolumn{7}{c|}{Training set size} \\\cline{2-8}
	Model & 1k & 3.5k & 10k & 25k & 35k\textsuperscript{*}&50k&110k\\\hline
	DTNN\textsuperscript{\cite{schutt2017quantum}}&-&-&1.2&1.0&-&0.94&-\\
	BAML\textsuperscript{\cite{huang2016understanding}} &-&-&2.4&-&-&-&-\\
	$\Delta^{B3LYP}_{MP7}-ML$\textsuperscript{\cite{ramakrishnan2015machine}} &4.8&-&3.0&-&-&-&-\\
	MPNN\textsuperscript{\cite{gilmer2017neural}}&-&-&-&-&-&-&\alert{0.39}\\
	HDAD\textsuperscript{\cite{faber2017fast}}&-&-&-&-&1.0&-&0.58\textsuperscript{*}\\
	HIP-NN\textsuperscript{\cite{lubbers2017hierarchical}}&-&-&-&-&-&\bf0.35&\bf0.26\\
	SchNet\textsuperscript{\cite{schutt2017moleculenet}}&-&-&-&-&-&0.59&0.31\\
	aSLATM$^{\cite{Huang2017opt}}$ &\bf1.8&\bf0.98$^{"}$&-&-&-&-&-\\\hline
	MTM$^\dagger$ &\bf1.8&\bf1.0$^{+}$&\bf0.86&\bf0.63&-&0.41&-\\\hline
\end{tabular}
\caption{Comparison of the MAE of prediction of atomization energy (kcal/mol) by different models for different training set sizes.
\\
{\small $^\dagger$\emph{this work}\hfill\strut}
\\
{\small*\emph{As estimated from the graphs in Ref.\ \onlinecite{faber2017fast}.\hfill\strut}}
\\
{\small$+$\emph{The error can be decreased by further training, see the text.}\hfill\strut}
\\
{\small $^"$\emph{as estimated from the graph in Ref. \onlinecite{Huang2017opt}}\hfill\strut}
}
\label{tab:gdb9comp}
\end{table}

\subsection*{Active learning}

\begin{figure}
	\centering
	\captionsetup[subfigure]{justification=centering}
	\subfigure[]{\includegraphics[scale=.60]{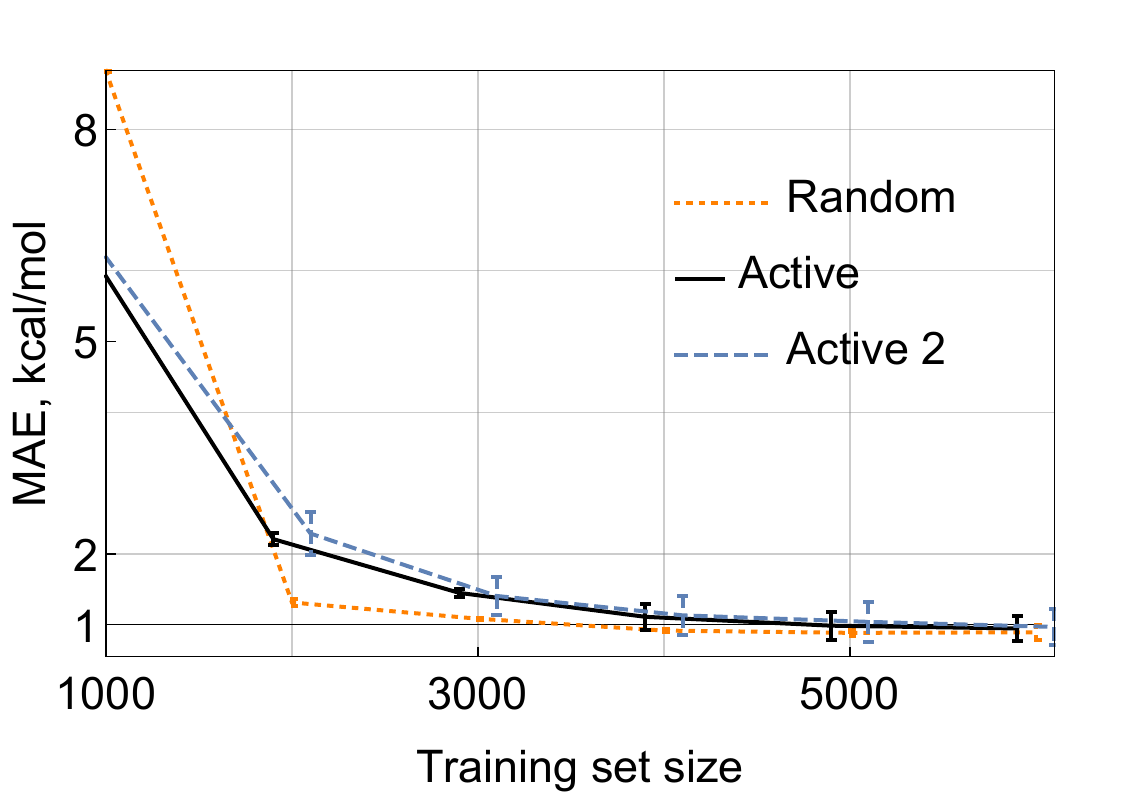}}
	\hfill
	\subfigure[]{\includegraphics[scale=.60]{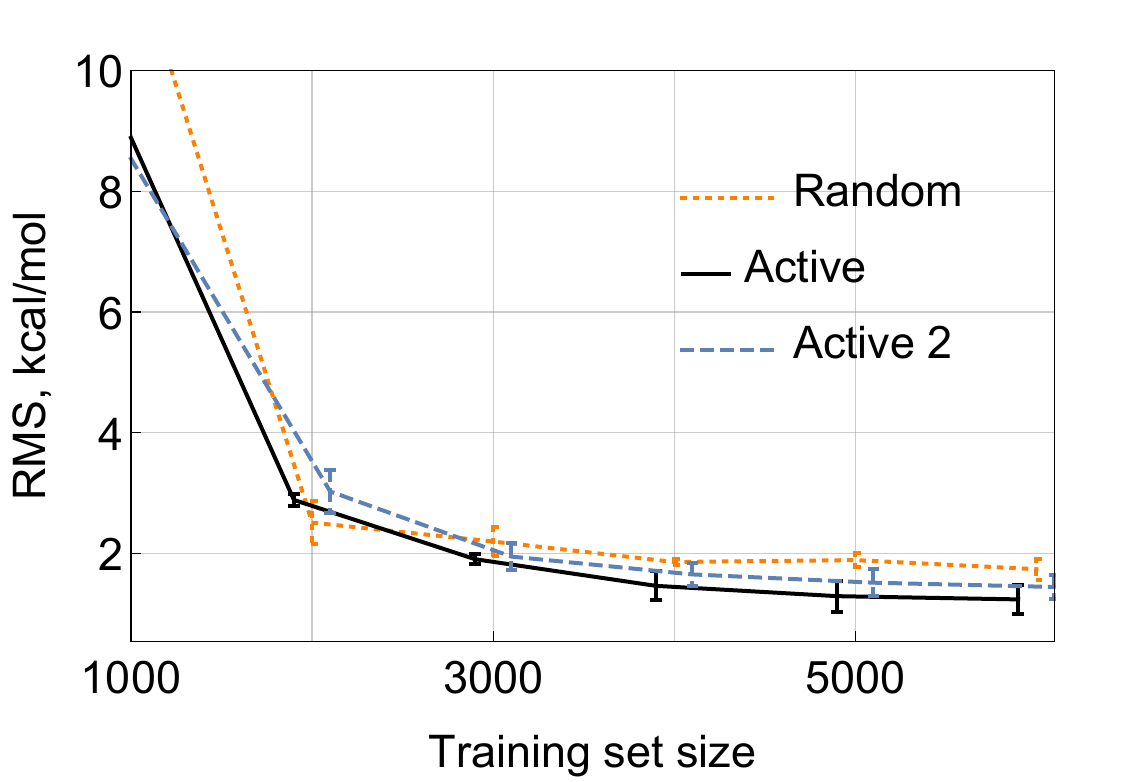}}
	\hfill
	\subfigure[]{\includegraphics[scale=.60]{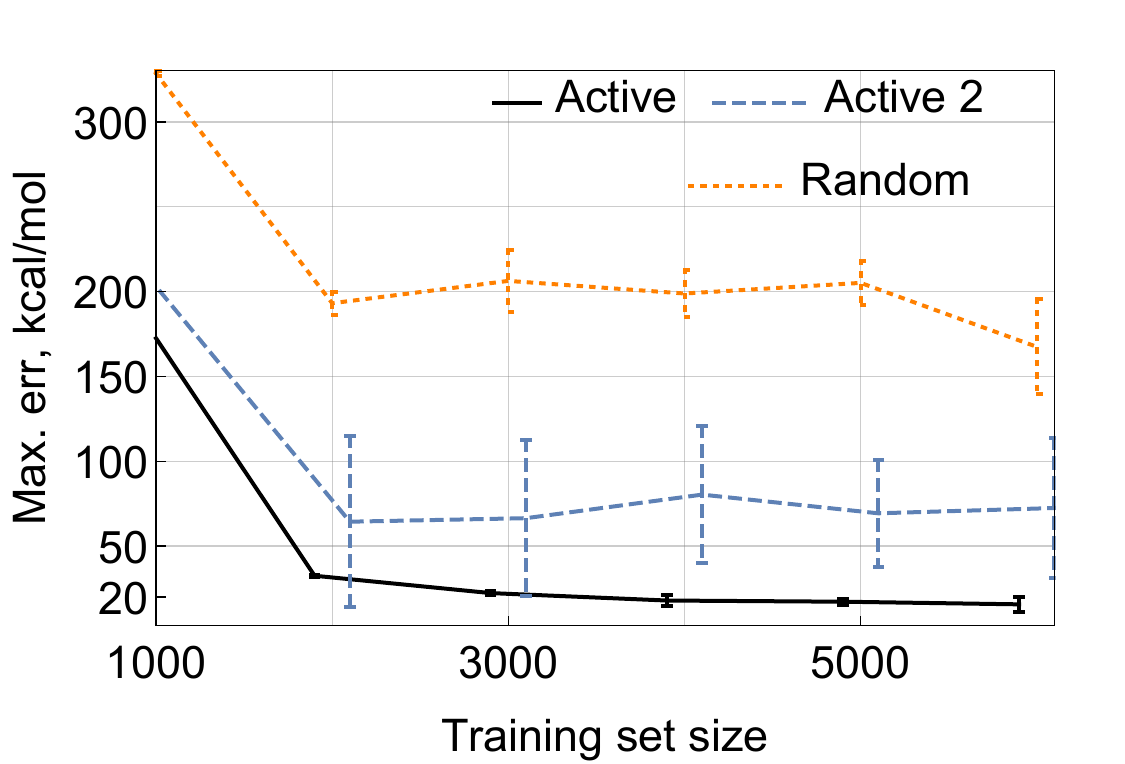}}
	\caption {Active and random selection of molecules: dependencies of MAE (a), RMSE (b) and maximal error (c) on the training set size for the MTM$_{24}$ model.}
	\label{fig:active}
\end{figure}

While the MAE of MTMs trained on random training sets is small, the corresponding maximal error is of the order of 100 kcal/mol, resulting in the outliers for which the atomization energy prediction is too inaccurate.
To get rid of the outliers we have applied the active learning algorithm as described in Section \ref{sec:al}. 
We started with a random training set of 1k molecules and it takes about 20 iterations to reach the training set size of 6k molecules. At each iteration, our training procedure takes an amount of time proportional to the training set size, while the selection always takes nearly constant time, approximately the same as required to train a model on a training set with 1k molecules.

Figure \ref{fig:active} shows the graphs of the MAE, RMSE and maximal absolute errors depending on the training set size, comparing random and active selection.
The ``Active'' curve corresponds to the scenario of actively selecting molecules from the entire set of molecules and measuring the error on the remaining set.
	In the ``Active 2'' scenario we instead separate out a validation set of 30k molecules on which we measure error, while selecting and learning from the remaining 100k molecules.
	The error bars correspond to the 95\% confidence interval as measured on three independent runs, in each of which the initial training set of 1000 molecules and the validation set in the ``Active 2'' scenario were random.
We have used the MTM$_{24}$ model. It has about 2000 parameters to fit, which explains why the error on the Figure \ref{fig:active} exhibits \textit{overfitting} when trained on less than 2000 molecules.
As stated in Section \ref{sec:alg}, we could improve our results for such small training set sizes by introducing regularization (which is otherwise not required).

In the active learning approach (see Section \ref{sec:al}), the training set tends to be as diverse as possible in the sense of spanning the largest volume of configurational (or molecular) space.
The most extrapolative molecules lie on the boundaries of this volume and if the amount of molecules is less than or close to the number of model parameters we would not have sufficient amount of training samples in-between the boundaries.
With the random selection we cover the configurational space more evenly, which results in lower RMS errors.
From the part of the graph starting from 4000 training samples (this is twice the number of model parameters, as we suggested as a rule of thumb in Section \ref{sec:alg}) we can see that there is no improvement in MAE (though it is rather close to the random-sampling MAE), but the RMS and maximal errors are lower that for the random sampling.
 
From Figure \ref{fig:active} it can be seen that the maximal error is much less when we are allowed to add any possible molecule to the training set (the ``Active'' scenario) as compared to the ``Active 2'' scenario, however, the error in the ``Active 2'' scenario is still smaller than that in random sampling.
This indicates that the active learning have two mechanisms of decreasing the error:
the actively chosen molecules represent better the unusual molecules in the validation set of the ``Active 2'' scenario, but if we are allowed to select also those unusual molecules for training, the error further drops.
We argue that the latter can be useful in those applications where the region of interest in the chemical space is fixed \textit{a priori}.
As an example of such application, in [\onlinecite{nyshadham2017computational}] the authors found six candidates of Co superalloys from about 2k \textit{a priori} identified potential chemical compositions.

%As we write in Section \ref{sec:al-scheme}, at the first stage of active learning we train our model on some randomly chosen molecules. The graphs on Figure \ref{fig:active} start from the training set size 1000, but actually we started our active learning procedure with a training set consisting of only 100 molecules.
%Note, that the MTM$_{24}$ has more than 1000 of parameters, therefore at this stage overfitting will for sure take place, worsening the predictions. 
%However, for the sake of comparison we wanted out training set to be enough ``actively picked'' by reaching the size of 1000.
%Still we note, that for best performance one should start the active learning procedure with a training set consisting of at least 1000 molecules.
%After almost each iteration (Section \ref{sec:al-scheme}) of molecules selection and model retraining, the maximal error drops, as the molecules with the most poorly reproduced enthalpies move to the training set, thus leaving the validation set. At the point, when we already picked 10k configurations, the maximal error on training set was about 9 kcal/mol, which is 50\% higher that on the validation set. 

\begin{figure}
	\centering
	\captionsetup[subfigure]{justification=centering}
	\subfigure[random choice of training samples]{\includegraphics[width=.45\textwidth]{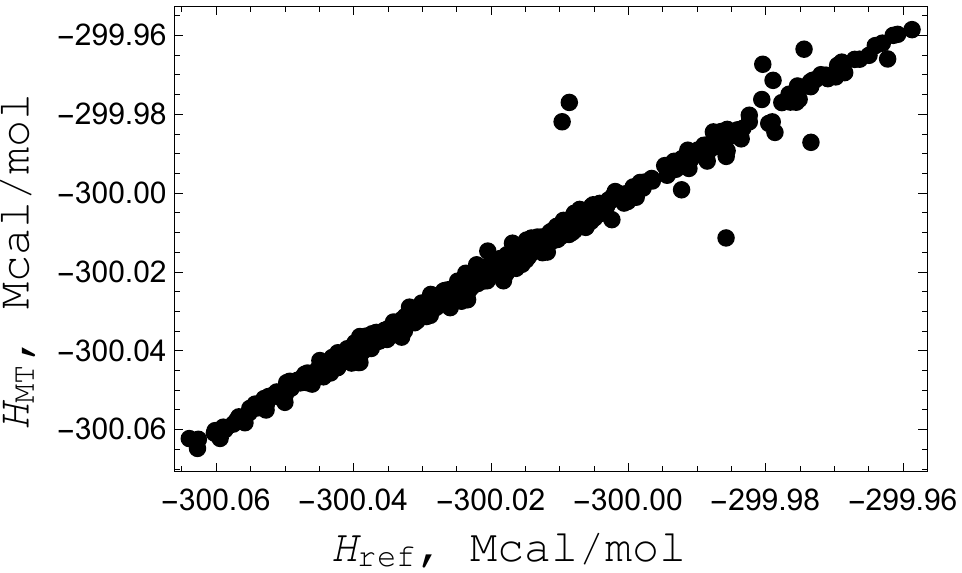}}
	\hfill
	\subfigure[active choice of training samples]{\includegraphics[width=.45\textwidth]{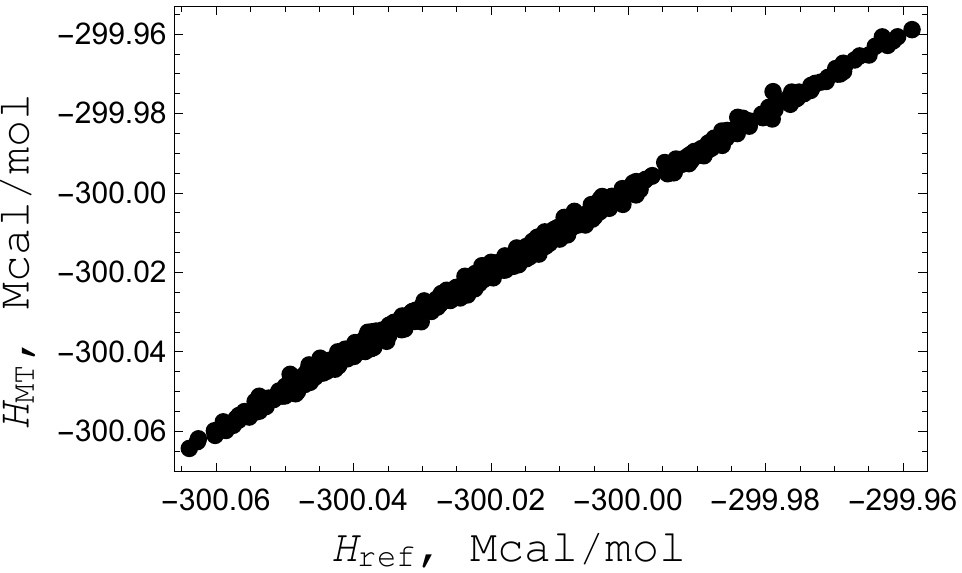}}
	\caption {Scatter plot of enthalpy predicted by machine learning, $H_{\rm MT}$, versus the reference enthalpy $H_{\rm ref}$, with (a) random and (b) active choice of training samples. For the illustration purposes we have plotted a small area from the whole database, where the errors are significant.}
	\label{fig:active_random_errs}
\end{figure}

To better understand the impact of active learning, on Figure \ref{fig:active_random_errs} we plotted the true and predicted enthalpies for the models trained on randomly and actively chosen training sets of 10k molecules.
The plot is focused on a small region of enthalpies to show the scale of the error.
As can be seen, active learning makes the error small uniformly over all the samples, while the random choice of the training set results in outliers for which the error is large.

\begin{figure}
	\centering
	\captionsetup[subfigure]{justification=centering}
	\includegraphics[scale=.5]{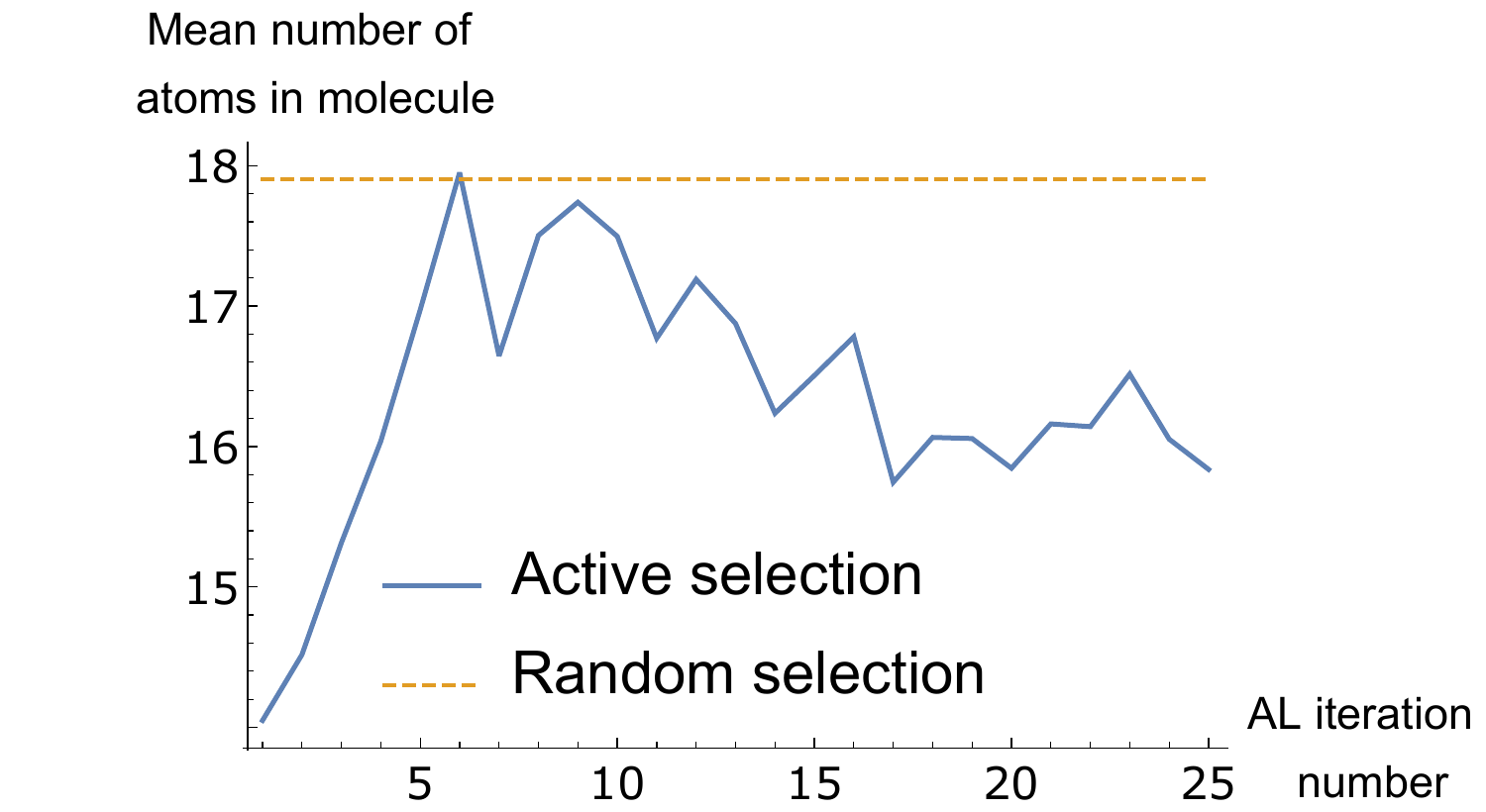}
	\hfill
	\caption {Mean numbers of atoms in molecule for active and random selection of configurations. The MTM$_{24}$ was used for this experiment. }
	\label{fig:mean_sizes}
\end{figure}

We investigated the sizes of the molecules entering the training set at each iteration of selection (see Section \ref{sec:al-scheme}) and compared them to the average molecule size among the QM9 database. The results are shown on Figure \ref{fig:mean_sizes}.
From this graphs it can be seen that the algorithm tries to select molecules with sizes lower than average.
This could indicate that some small molecules contain representative atomic neighborhoods which occur in a many bigger molecules and the active learning algorithm detects and selects such small molecules.
This result is in correspondence with the observations of Huang et al.\ \cite{Huang2017opt}, where the authors state that an accurate model can be obtained by training on a small amount of the most relevant atomic environments (neighborhoods). 

\subsection*{Fitting the QM7 database}
 
Another common benchmark database we used consists of 7.2k small organic molecules with up to seven heavy atoms (C,N,O,S) saturated by H, and is referred to as QM7  \cite{montavon2013machine}.
We have chosen the four properties to fit: atomization energy, polarizability, and HOMO and LUMO levels.
The atomization energy and polarizability are extensive quantities, and the other two are intensive quantities.
The training set consisted of 5k randomly chosen molecules while the remaining 2200 were used for validation. In Table \ref{tab:compar} we compare the accuracy of the MTMs to the other state-of-the-art methods. Our results are averaged over three independent runs.
We see that the local properties, enthalpy and polarizability, are fitted with the same or higher accuracy as by the existing methods, while predictions of the HOMO and LUMO levels have a 50\% larger error than the state of the art.

To address this issue, we have applied the nonlocal modification of our algorithm, nlMTM, as given by \eqref{eq:nonlocal-model}.
From Table \ref{tab:compar} we see that accounting for nonlocality improves the error of HOMO and LUMO to the state-of-the-art accuracy.
As, the nonlocal scheme is, essentially, a three-layer model, it was found to suffer from overfitting similarly to the deep neural networks; therefore we used the early stopping technique for training the nlMTM.
To that end, we used 1100 samples for estimating the error during training and measured the error on the remaining 1100 samples.

%The best MAEs for HOMO and LUMO in Table \ref{tab:compar} we have obtained with the non-local version of our model (\ref{eq:noloc}). 
%While the local model (\ref{eq:part}) provided MAE about 0.16 eV for both HOMO and LUMO, we decided that it is due to significant non-locality of the mentioned properties, which should be treated with (\ref{eq:noloc}).
%We improved the accuracy by extending the functional form to capture the non-localities of the atomic systems. With such an approach we get rid of locality principle, when the function $F(x)$ from \eqref{eq:part} is composed of individual contributions of atomic environments. While non-local approach allows us to drop the MAE for the HOMO and LUMO by 30\%, the training of \eqref{eq:noloc} is a hard task. 

 \begin{table} 
 	\centering
 \begin{tabular}{| l |c |c|c|c|} 
 	\hline			
 	Model &$E$&$\alpha$& HOMO& LUMO\\
 	     &(kcal/mol)&(\AA$^3$)& (eV)& (eV)\\
 	\hline	
 	BAML\textsuperscript{\cite{huang2016understanding}} & 1.15 & 0.07&{\bf 0.10}&{\bf 0.11} \\
 	\hline  
 	SOAP\textsuperscript{\cite{de2016comparing}} & 0.92 & 0.05 &0.12&0.12\\
 	\hline  
 	DTNN\textsuperscript{\cite{schutt2017quantum}}&1.04& -&-&-\\
 	\hline  
 	MBTR\textsuperscript{\cite{huo2017unified}}& 0.60 & {\bf 0.04} &-&-\\
 	\hline  
 	MTM$^\dagger$ & {\bf 0.52} & {\bf 0.04} &0.16&0.16\\
 	\hline  
 	nlMTM$^\dagger$ & - & - &0.11&{\bf 0.11}\\
 	\hline  
 \end{tabular}
	\caption{Mean average errors of predicting the atomization energy $E$, polarizability $\alpha$, and HOMO and LUMO levels, committed by different models on the QM7 dataset.
	\\
	{\small$^\dagger$\emph{this work}}\hfill\strut
	}
	\label{tab:compar}
\end{table}

\section{Conclusions}\label{sec:conclusion}
 
We have introduced the moment tensor model (MTM) for prediction of the molecular properties based on the contributions of local atomic environments represented by the moment tensor descriptors.
The accuracy of the proposed model is comparable or better than that of the state-of-the-art algorithms.
The proposed model outperforms the existing algorithms by how fast it reaches the chemical accuracy on a database of 130k molecules.
We attribute this to the provable completeness of our moment tensor descriptors of local environments.
It should be emphasized that although our descriptors \eqref{eq:moments} are, essentially, two-body descriptors, their completeness and, in particular, angular dependence comes from their tensorial structure.
%The locality principle has proved its effectiveness as the model learns from the atomic environments and afterwards is able to predict properties of the complex molecules based on their decomposition into such environments. This results in remarkably faster learning rate (accuracy vs training set size) comparing to other models.

In addition, we have proposed an active learning algorithm that significantly reduces the maximal error (in other words, the error for the outliers).
The algorithm effectively selects the molecules most different from those already in the training set and adds these molecules to the training set.

%\commentas{TODO: consider giving an interpolation/extrapolation interpretation, then discuss it here.}
%which prevents extrapolation in terms of geometry and chemical composition - the reason for outliers appearance as such. 

%Summing everything up, we conclude that the proposed model possesses certain qualities which makes it a suitable choice for the solution of such problem as molecular properties prediction. 
  
\section*{Acknowledgements}

We thank Matthias Rupp and Anatole von Lilienfeld for fruitful discussions of our results. The work was supported by the Skoltech NGP Program No.\ 2016-7/NGP (a Skoltech-MIT joint project).
This work was performed, in part, by A.S. during the Fall 2016 long program at the Institute of Pure and Applied Mathematics, UCLA; and, in part, at the Center for Integrated Nanotechnologies, an Office of Science User Facility operated for the U.S. Department of Energy (DOE) Office of Science by Los Alamos National Laboratory (Contract DE-AC52-06NA25396) and Sandia National Laboratories (Contract DE-NA-0003525).

\bibliography{paper}	

%merlin.mbs aipnum4-1.bst 2010-07-25 4.21a (PWD, AO, DPC) hacked
%Control: key (0)
%Control: author (8) initials jnrlst
%Control: editor formatted (1) identically to author
%Control: production of article title (0) allowed
%Control: page (1) range
%Control: year (1) truncated
%Control: production of eprint (0) enabled
\begin{thebibliography}{21}%
\makeatletter
\providecommand \@ifxundefined [1]{%
 \@ifx{#1\undefined}
}%
\providecommand \@ifnum [1]{%
 \ifnum #1\expandafter \@firstoftwo
 \else \expandafter \@secondoftwo
 \fi
}%
\providecommand \@ifx [1]{%
 \ifx #1\expandafter \@firstoftwo
 \else \expandafter \@secondoftwo
 \fi
}%
\providecommand \natexlab [1]{#1}%
\providecommand \enquote  [1]{``#1''}%
\providecommand \bibnamefont  [1]{#1}%
\providecommand \bibfnamefont [1]{#1}%
\providecommand \citenamefont [1]{#1}%
\providecommand \href@noop [0]{\@secondoftwo}%
\providecommand \href [0]{\begingroup \@sanitize@url \@href}%
\providecommand \@href[1]{\@@startlink{#1}\@@href}%
\providecommand \@@href[1]{\endgroup#1\@@endlink}%
\providecommand \@sanitize@url [0]{\catcode `\\12\catcode `\$12\catcode
  `\&12\catcode `\#12\catcode `\^12\catcode `\_12\catcode `\%12\relax}%
\providecommand \@@startlink[1]{}%
\providecommand \@@endlink[0]{}%
\providecommand \url  [0]{\begingroup\@sanitize@url \@url }%
\providecommand \@url [1]{\endgroup\@href {#1}{\urlprefix }}%
\providecommand \urlprefix  [0]{URL }%
\providecommand \Eprint [0]{\href }%
\providecommand \doibase [0]{http://dx.doi.org/}%
\providecommand \selectlanguage [0]{\@gobble}%
\providecommand \bibinfo  [0]{\@secondoftwo}%
\providecommand \bibfield  [0]{\@secondoftwo}%
\providecommand \translation [1]{[#1]}%
\providecommand \BibitemOpen [0]{}%
\providecommand \bibitemStop [0]{}%
\providecommand \bibitemNoStop [0]{.\EOS\space}%
\providecommand \EOS [0]{\spacefactor3000\relax}%
\providecommand \BibitemShut  [1]{\csname bibitem#1\endcsname}%
\let\auto@bib@innerbib\@empty
%</preamble>
\bibitem [{\citenamefont {Browning}\ \emph {et~al.}(2017)\citenamefont
  {Browning}, \citenamefont {Ramakrishnan}, \citenamefont {von Lilienfeld},\
  and\ \citenamefont {Roethlisberger}}]{browning2017genetic}%
  \BibitemOpen
  \bibfield  {author} {\bibinfo {author} {\bibfnamefont {N.~J.}\ \bibnamefont
  {Browning}}, \bibinfo {author} {\bibfnamefont {R.}~\bibnamefont
  {Ramakrishnan}}, \bibinfo {author} {\bibfnamefont {O.~A.}\ \bibnamefont {von
  Lilienfeld}}, \ and\ \bibinfo {author} {\bibfnamefont {U.}~\bibnamefont
  {Roethlisberger}},\ }\bibfield  {title} {\enquote {\bibinfo {title} {Genetic
  optimization of training sets for improved machine learning models of
  molecular properties},}\ }\href@noop {} {\bibfield  {journal} {\bibinfo
  {journal} {The Journal of Physical Chemistry Letters}\ }\textbf {\bibinfo
  {volume} {8}},\ \bibinfo {pages} {1351--1359} (\bibinfo {year}
  {2017})}\BibitemShut {NoStop}%
\bibitem [{\citenamefont {Hansen}\ \emph {et~al.}(2015)\citenamefont {Hansen},
  \citenamefont {Biegler}, \citenamefont {Ramakrishnan}, \citenamefont
  {Pronobis}, \citenamefont {Von~Lilienfeld}, \citenamefont {Müller},\ and\
  \citenamefont {Tkatchenko}}]{hansen2015machine}%
  \BibitemOpen
  \bibfield  {author} {\bibinfo {author} {\bibfnamefont {K.}~\bibnamefont
  {Hansen}}, \bibinfo {author} {\bibfnamefont {F.}~\bibnamefont {Biegler}},
  \bibinfo {author} {\bibfnamefont {R.}~\bibnamefont {Ramakrishnan}}, \bibinfo
  {author} {\bibfnamefont {W.}~\bibnamefont {Pronobis}}, \bibinfo {author}
  {\bibfnamefont {O.~A.}\ \bibnamefont {Von~Lilienfeld}}, \bibinfo {author}
  {\bibfnamefont {K.-R.}\ \bibnamefont {Müller}}, \ and\ \bibinfo {author}
  {\bibfnamefont {A.}~\bibnamefont {Tkatchenko}},\ }\bibfield  {title}
  {\enquote {\bibinfo {title} {Machine learning predictions of molecular
  properties: Accurate many-body potentials and nonlocality in chemical
  space},}\ }\href@noop {} {\bibfield  {journal} {\bibinfo  {journal} {The
  journal of physical chemistry letters}\ }\textbf {\bibinfo {volume} {6}},\
  \bibinfo {pages} {2326--2331} (\bibinfo {year} {2015})}\BibitemShut {NoStop}%
\bibitem [{\citenamefont {Ramakrishnan}\ \emph {et~al.}(2014)\citenamefont
  {Ramakrishnan}, \citenamefont {Dral}, \citenamefont {Rupp},\ and\
  \citenamefont {Von~Lilienfeld}}]{ramakrishnan2014quantum}%
  \BibitemOpen
  \bibfield  {author} {\bibinfo {author} {\bibfnamefont {R.}~\bibnamefont
  {Ramakrishnan}}, \bibinfo {author} {\bibfnamefont {P.~O.}\ \bibnamefont
  {Dral}}, \bibinfo {author} {\bibfnamefont {M.}~\bibnamefont {Rupp}}, \ and\
  \bibinfo {author} {\bibfnamefont {O.~A.}\ \bibnamefont {Von~Lilienfeld}},\
  }\bibfield  {title} {\enquote {\bibinfo {title} {Quantum chemistry structures
  and properties of 134 kilo molecules},}\ }\href@noop {} {\bibfield  {journal}
  {\bibinfo  {journal} {Scientific data}\ }\textbf {\bibinfo {volume} {1}}
  (\bibinfo {year} {2014})}\BibitemShut {NoStop}%
\bibitem [{\citenamefont {Ramakrishnan}\ \emph {et~al.}(2015)\citenamefont
  {Ramakrishnan}, \citenamefont {Dral}, \citenamefont {Rupp},\ and\
  \citenamefont {von Lilienfeld}}]{ramakrishnan2015big}%
  \BibitemOpen
  \bibfield  {author} {\bibinfo {author} {\bibfnamefont {R.}~\bibnamefont
  {Ramakrishnan}}, \bibinfo {author} {\bibfnamefont {P.~O.}\ \bibnamefont
  {Dral}}, \bibinfo {author} {\bibfnamefont {M.}~\bibnamefont {Rupp}}, \ and\
  \bibinfo {author} {\bibfnamefont {O.~A.}\ \bibnamefont {von Lilienfeld}},\
  }\bibfield  {title} {\enquote {\bibinfo {title} {Big data meets quantum
  chemistry approximations: the $\delta$-machine learning approach},}\
  }\href@noop {} {\bibfield  {journal} {\bibinfo  {journal} {Journal of
  chemical theory and computation}\ }\textbf {\bibinfo {volume} {11}},\
  \bibinfo {pages} {2087--2096} (\bibinfo {year} {2015})}\BibitemShut {NoStop}%
\bibitem [{\citenamefont {Ramakrishnan}\ and\ \citenamefont {von
  Lilienfeld}(2015)}]{ramakrishnan2015machine}%
  \BibitemOpen
  \bibfield  {author} {\bibinfo {author} {\bibfnamefont {R.}~\bibnamefont
  {Ramakrishnan}}\ and\ \bibinfo {author} {\bibfnamefont {O.~A.}\ \bibnamefont
  {von Lilienfeld}},\ }\bibfield  {title} {\enquote {\bibinfo {title} {Machine
  learning, quantum mechanics, and chemical compound space},}\ }\href@noop {}
  {\bibfield  {journal} {\bibinfo  {journal} {arXiv preprint arXiv:1510.07512}\
  } (\bibinfo {year} {2015})}\BibitemShut {NoStop}%
\bibitem [{\citenamefont {Huang}\ and\ \citenamefont
  {Von~Lilienfeld}(2016)}]{huang2016understanding}%
  \BibitemOpen
  \bibfield  {author} {\bibinfo {author} {\bibfnamefont {B.}~\bibnamefont
  {Huang}}\ and\ \bibinfo {author} {\bibfnamefont {O.~A.}\ \bibnamefont
  {Von~Lilienfeld}},\ }\bibfield  {title} {\enquote {\bibinfo {title}
  {Communication: Understanding molecular representations in machine learning:
  The role of uniqueness and target similarity},}\ }\href@noop {} {\bibfield
  {journal} {\bibinfo  {journal} {Journal of Chemical Physics}\ }\textbf
  {\bibinfo {volume} {145}} (\bibinfo {year} {2016})}\BibitemShut {NoStop}%
\bibitem [{\citenamefont {Sch{\"u}tt}\ \emph
  {et~al.}(2017{\natexlab{a}})\citenamefont {Sch{\"u}tt}, \citenamefont
  {Arbabzadah}, \citenamefont {Chmiela}, \citenamefont {M{\"u}ller},\ and\
  \citenamefont {Tkatchenko}}]{schutt2017quantum}%
  \BibitemOpen
  \bibfield  {author} {\bibinfo {author} {\bibfnamefont {K.~T.}\ \bibnamefont
  {Sch{\"u}tt}}, \bibinfo {author} {\bibfnamefont {F.}~\bibnamefont
  {Arbabzadah}}, \bibinfo {author} {\bibfnamefont {S.}~\bibnamefont {Chmiela}},
  \bibinfo {author} {\bibfnamefont {K.~R.}\ \bibnamefont {M{\"u}ller}}, \ and\
  \bibinfo {author} {\bibfnamefont {A.}~\bibnamefont {Tkatchenko}},\ }\bibfield
   {title} {\enquote {\bibinfo {title} {Quantum-chemical insights from deep
  tensor neural networks},}\ }\href@noop {} {\bibfield  {journal} {\bibinfo
  {journal} {Nature communications}\ }\textbf {\bibinfo {volume} {8}},\
  \bibinfo {pages} {13890} (\bibinfo {year} {2017}{\natexlab{a}})}\BibitemShut
  {NoStop}%
\bibitem [{\citenamefont {Faber}\ \emph {et~al.}(2017)\citenamefont {Faber},
  \citenamefont {Hutchison}, \citenamefont {Huang}, \citenamefont {Gilmer},
  \citenamefont {Schoenholz}, \citenamefont {Dahl}, \citenamefont {Vinyals},
  \citenamefont {Kearnes}, \citenamefont {Riley},\ and\ \citenamefont {von
  Lilienfeld}}]{faber2017fast}%
  \BibitemOpen
  \bibfield  {author} {\bibinfo {author} {\bibfnamefont {F.~A.}\ \bibnamefont
  {Faber}}, \bibinfo {author} {\bibfnamefont {L.}~\bibnamefont {Hutchison}},
  \bibinfo {author} {\bibfnamefont {B.}~\bibnamefont {Huang}}, \bibinfo
  {author} {\bibfnamefont {J.}~\bibnamefont {Gilmer}}, \bibinfo {author}
  {\bibfnamefont {S.~S.}\ \bibnamefont {Schoenholz}}, \bibinfo {author}
  {\bibfnamefont {G.~E.}\ \bibnamefont {Dahl}}, \bibinfo {author}
  {\bibfnamefont {O.}~\bibnamefont {Vinyals}}, \bibinfo {author} {\bibfnamefont
  {S.}~\bibnamefont {Kearnes}}, \bibinfo {author} {\bibfnamefont {P.~F.}\
  \bibnamefont {Riley}}, \ and\ \bibinfo {author} {\bibfnamefont {O.~A.}\
  \bibnamefont {von Lilienfeld}},\ }\bibfield  {title} {\enquote {\bibinfo
  {title} {Fast machine learning models of electronic and energetic properties
  consistently reach approximation errors better than {DFT} accuracy},}\
  }\href@noop {} {\bibfield  {journal} {\bibinfo  {journal} {arXiv preprint
  arXiv:1702.05532}\ } (\bibinfo {year} {2017})}\BibitemShut {NoStop}%
\bibitem [{\citenamefont {Gilmer}\ \emph {et~al.}(2017)\citenamefont {Gilmer},
  \citenamefont {Schoenholz}, \citenamefont {Riley}, \citenamefont {Vinyals},\
  and\ \citenamefont {Dahl}}]{gilmer2017neural}%
  \BibitemOpen
  \bibfield  {author} {\bibinfo {author} {\bibfnamefont {J.}~\bibnamefont
  {Gilmer}}, \bibinfo {author} {\bibfnamefont {S.~S.}\ \bibnamefont
  {Schoenholz}}, \bibinfo {author} {\bibfnamefont {P.~F.}\ \bibnamefont
  {Riley}}, \bibinfo {author} {\bibfnamefont {O.}~\bibnamefont {Vinyals}}, \
  and\ \bibinfo {author} {\bibfnamefont {G.~E.}\ \bibnamefont {Dahl}},\
  }\bibfield  {title} {\enquote {\bibinfo {title} {Neural message passing for
  quantum chemistry},}\ }\href@noop {} {\bibfield  {journal} {\bibinfo
  {journal} {arXiv preprint arXiv:1704.01212}\ } (\bibinfo {year}
  {2017})}\BibitemShut {NoStop}%
\bibitem [{\citenamefont {Huo}\ and\ \citenamefont
  {Rupp}(2017)}]{huo2017unified}%
  \BibitemOpen
  \bibfield  {author} {\bibinfo {author} {\bibfnamefont {H.}~\bibnamefont
  {Huo}}\ and\ \bibinfo {author} {\bibfnamefont {M.}~\bibnamefont {Rupp}},\
  }\bibfield  {title} {\enquote {\bibinfo {title} {Unified representation for
  machine learning of molecules and crystals},}\ }\href@noop {} {\bibfield
  {journal} {\bibinfo  {journal} {arXiv preprint arXiv:1704.06439}\ } (\bibinfo
  {year} {2017})}\BibitemShut {NoStop}%
\bibitem [{\citenamefont {Rupp}\ \emph {et~al.}(2012)\citenamefont {Rupp},
  \citenamefont {Tkatchenko}, \citenamefont {M{\"u}ller},\ and\ \citenamefont
  {Von~Lilienfeld}}]{Rupp2012}%
  \BibitemOpen
  \bibfield  {author} {\bibinfo {author} {\bibfnamefont {M.}~\bibnamefont
  {Rupp}}, \bibinfo {author} {\bibfnamefont {A.}~\bibnamefont {Tkatchenko}},
  \bibinfo {author} {\bibfnamefont {K.-R.}\ \bibnamefont {M{\"u}ller}}, \ and\
  \bibinfo {author} {\bibfnamefont {O.~A.}\ \bibnamefont {Von~Lilienfeld}},\
  }\bibfield  {title} {\enquote {\bibinfo {title} {Fast and accurate modeling
  of molecular atomization energies with machine learning},}\ }\href@noop {}
  {\bibfield  {journal} {\bibinfo  {journal} {Physical review letters}\
  }\textbf {\bibinfo {volume} {108}},\ \bibinfo {pages} {058301} (\bibinfo
  {year} {2012})}\BibitemShut {NoStop}%
\bibitem [{\citenamefont {Rupp}, \citenamefont {Ramakrishnan},\ and\
  \citenamefont {von Lilienfeld}(2015)}]{rupp2015machine}%
  \BibitemOpen
  \bibfield  {author} {\bibinfo {author} {\bibfnamefont {M.}~\bibnamefont
  {Rupp}}, \bibinfo {author} {\bibfnamefont {R.}~\bibnamefont {Ramakrishnan}},
  \ and\ \bibinfo {author} {\bibfnamefont {O.~A.}\ \bibnamefont {von
  Lilienfeld}},\ }\bibfield  {title} {\enquote {\bibinfo {title} {Machine
  learning for quantum mechanical properties of atoms in molecules},}\
  }\href@noop {} {\bibfield  {journal} {\bibinfo  {journal} {The Journal of
  Physical Chemistry Letters}\ }\textbf {\bibinfo {volume} {6}},\ \bibinfo
  {pages} {3309--3313} (\bibinfo {year} {2015})}\BibitemShut {NoStop}%
\bibitem [{\citenamefont {De}\ \emph {et~al.}(2016)\citenamefont {De},
  \citenamefont {Bart{\'o}k}, \citenamefont {Cs{\'a}nyi},\ and\ \citenamefont
  {Ceriotti}}]{de2016comparing}%
  \BibitemOpen
  \bibfield  {author} {\bibinfo {author} {\bibfnamefont {S.}~\bibnamefont
  {De}}, \bibinfo {author} {\bibfnamefont {A.~P.}\ \bibnamefont {Bart{\'o}k}},
  \bibinfo {author} {\bibfnamefont {G.}~\bibnamefont {Cs{\'a}nyi}}, \ and\
  \bibinfo {author} {\bibfnamefont {M.}~\bibnamefont {Ceriotti}},\ }\bibfield
  {title} {\enquote {\bibinfo {title} {Comparing molecules and solids across
  structural and alchemical space},}\ }\href@noop {} {\bibfield  {journal}
  {\bibinfo  {journal} {Physical Chemistry Chemical Physics}\ }\textbf
  {\bibinfo {volume} {18}},\ \bibinfo {pages} {13754--13769} (\bibinfo {year}
  {2016})}\BibitemShut {NoStop}%
\bibitem [{\citenamefont {Lubbers}, \citenamefont {Smith},\ and\ \citenamefont
  {Barros}(2017)}]{lubbers2017hierarchical}%
  \BibitemOpen
  \bibfield  {author} {\bibinfo {author} {\bibfnamefont {N.}~\bibnamefont
  {Lubbers}}, \bibinfo {author} {\bibfnamefont {J.~S.}\ \bibnamefont {Smith}},
  \ and\ \bibinfo {author} {\bibfnamefont {K.}~\bibnamefont {Barros}},\
  }\bibfield  {title} {\enquote {\bibinfo {title} {Hierarchical modeling of
  molecular energies using a deep neural network},}\ }\href@noop {} {\bibfield
  {journal} {\bibinfo  {journal} {arXiv preprint arXiv:1710.00017}\ } (\bibinfo
  {year} {2017})}\BibitemShut {NoStop}%
\bibitem [{\citenamefont {Sch{\"u}tt}\ \emph
  {et~al.}(2017{\natexlab{b}})\citenamefont {Sch{\"u}tt}, \citenamefont
  {Kindermans}, \citenamefont {Felix}, \citenamefont {Chmiela}, \citenamefont
  {Tkatchenko},\ and\ \citenamefont {M{\"u}ller}}]{schutt2017moleculenet}%
  \BibitemOpen
  \bibfield  {author} {\bibinfo {author} {\bibfnamefont {K.}~\bibnamefont
  {Sch{\"u}tt}}, \bibinfo {author} {\bibfnamefont {P.-J.}\ \bibnamefont
  {Kindermans}}, \bibinfo {author} {\bibfnamefont {H.~E.~S.}\ \bibnamefont
  {Felix}}, \bibinfo {author} {\bibfnamefont {S.}~\bibnamefont {Chmiela}},
  \bibinfo {author} {\bibfnamefont {A.}~\bibnamefont {Tkatchenko}}, \ and\
  \bibinfo {author} {\bibfnamefont {K.-R.}\ \bibnamefont {M{\"u}ller}},\
  }\bibfield  {title} {\enquote {\bibinfo {title} {Moleculenet: A
  continuous-filter convolutional neural network for modeling quantum
  interactions},}\ }in\ \href@noop {} {\emph {\bibinfo {booktitle} {Advances in
  Neural Information Processing Systems}}}\ (\bibinfo {year} {2017})\ pp.\
  \bibinfo {pages} {992--1002}\BibitemShut {NoStop}%
\bibitem [{\citenamefont {Shapeev}(2016)}]{shapeev2016moment}%
  \BibitemOpen
  \bibfield  {author} {\bibinfo {author} {\bibfnamefont {A.~V.}\ \bibnamefont
  {Shapeev}},\ }\bibfield  {title} {\enquote {\bibinfo {title} {Moment tensor
  potentials: a class of systematically improvable interatomic potentials},}\
  }\href@noop {} {\bibfield  {journal} {\bibinfo  {journal} {Multiscale
  Modeling \& Simulation}\ }\textbf {\bibinfo {volume} {14}},\ \bibinfo {pages}
  {1153--1173} (\bibinfo {year} {2016})}\BibitemShut {NoStop}%
\bibitem [{\citenamefont {Huang}\ and\ \citenamefont {von
  Lilienfeld}(2017)}]{Huang2017opt}%
  \BibitemOpen
  \bibfield  {author} {\bibinfo {author} {\bibfnamefont {B.}~\bibnamefont
  {Huang}}\ and\ \bibinfo {author} {\bibfnamefont {O.~A.}\ \bibnamefont {von
  Lilienfeld}},\ }\bibfield  {title} {\enquote {\bibinfo {title} {The "dna" of
  chemistry: Scalable quantum machine learning with "amons"},}\ }\href@noop {}
  {\bibfield  {journal} {\bibinfo  {journal} {arXiv preprint arXiv:1707.04146}\
  } (\bibinfo {year} {2017})}\BibitemShut {NoStop}%
\bibitem [{\citenamefont {Podryabinkin}\ and\ \citenamefont
  {Shapeev}(2016)}]{podryabinkin2016active}%
  \BibitemOpen
  \bibfield  {author} {\bibinfo {author} {\bibfnamefont {E.~V.}\ \bibnamefont
  {Podryabinkin}}\ and\ \bibinfo {author} {\bibfnamefont {A.~V.}\ \bibnamefont
  {Shapeev}},\ }\bibfield  {title} {\enquote {\bibinfo {title} {Active learning
  of linear interatomic potentials},}\ }\href@noop {} {\bibfield  {journal}
  {\bibinfo  {journal} {arXiv preprint arXiv:1611.09346}\ } (\bibinfo {year}
  {2016})}\BibitemShut {NoStop}%
\bibitem [{\citenamefont {Goreinov}\ \emph {et~al.}(2010)\citenamefont
  {Goreinov}, \citenamefont {Oseledets}, \citenamefont {Savostyanov},
  \citenamefont {Tyrtyshnikov},\ and\ \citenamefont
  {Zamarashkin}}]{oseledets2010how-to1770566}%
  \BibitemOpen
  \bibfield  {author} {\bibinfo {author} {\bibfnamefont {S.}~\bibnamefont
  {Goreinov}}, \bibinfo {author} {\bibfnamefont {I.}~\bibnamefont {Oseledets}},
  \bibinfo {author} {\bibfnamefont {D.}~\bibnamefont {Savostyanov}}, \bibinfo
  {author} {\bibfnamefont {E.}~\bibnamefont {Tyrtyshnikov}}, \ and\ \bibinfo
  {author} {\bibfnamefont {N.}~\bibnamefont {Zamarashkin}},\ }\bibfield
  {title} {\enquote {\bibinfo {title} {How to find a good submatrix},}\ }in\
  \href@noop {} {\emph {\bibinfo {booktitle} {Matrix Methods: Theory,
  Algorithms, Applications}}}\ (\bibinfo  {publisher} {Word Scientific},\
  \bibinfo {year} {2010})\ pp.\ \bibinfo {pages} {247--256}\BibitemShut
  {NoStop}%
\bibitem [{\citenamefont {Ruddigkeit}\ \emph {et~al.}(2012)\citenamefont
  {Ruddigkeit}, \citenamefont {Van~Deursen}, \citenamefont {Blum},\ and\
  \citenamefont {Reymond}}]{ruddigkeit2012enumeration}%
  \BibitemOpen
  \bibfield  {author} {\bibinfo {author} {\bibfnamefont {L.}~\bibnamefont
  {Ruddigkeit}}, \bibinfo {author} {\bibfnamefont {R.}~\bibnamefont
  {Van~Deursen}}, \bibinfo {author} {\bibfnamefont {L.~C.}\ \bibnamefont
  {Blum}}, \ and\ \bibinfo {author} {\bibfnamefont {J.-L.}\ \bibnamefont
  {Reymond}},\ }\bibfield  {title} {\enquote {\bibinfo {title} {Enumeration of
  166 billion organic small molecules in the chemical universe database
  gdb-17},}\ }\href@noop {} {\bibfield  {journal} {\bibinfo  {journal} {Journal
  of chemical information and modeling}\ }\textbf {\bibinfo {volume} {52}},\
  \bibinfo {pages} {2864--2875} (\bibinfo {year} {2012})}\BibitemShut {NoStop}%
\bibitem [{\citenamefont {Nyshadham}\ \emph {et~al.}(2017)\citenamefont
  {Nyshadham}, \citenamefont {Oses}, \citenamefont {Hansen}, \citenamefont
  {Takeuchi}, \citenamefont {Curtarolo},\ and\ \citenamefont
  {Hart}}]{nyshadham2017computational}%
  \BibitemOpen
  \bibfield  {author} {\bibinfo {author} {\bibfnamefont {C.}~\bibnamefont
  {Nyshadham}}, \bibinfo {author} {\bibfnamefont {C.}~\bibnamefont {Oses}},
  \bibinfo {author} {\bibfnamefont {J.~E.}\ \bibnamefont {Hansen}}, \bibinfo
  {author} {\bibfnamefont {I.}~\bibnamefont {Takeuchi}}, \bibinfo {author}
  {\bibfnamefont {S.}~\bibnamefont {Curtarolo}}, \ and\ \bibinfo {author}
  {\bibfnamefont {G.~L.}\ \bibnamefont {Hart}},\ }\bibfield  {title} {\enquote
  {\bibinfo {title} {A computational high-throughput search for new ternary
  superalloys},}\ }\href@noop {} {\bibfield  {journal} {\bibinfo  {journal}
  {Acta Materialia}\ }\textbf {\bibinfo {volume} {122}},\ \bibinfo {pages}
  {438--447} (\bibinfo {year} {2017})}\BibitemShut {NoStop}%
\end{thebibliography}%

\end{document}